\documentclass[useAMS]{mn2e}
\usepackage{times}
\usepackage{graphicx}

\title[{\it Chandra\/} monitoring observations of NGC 5204 X-1]{{\it
Chandra\/} monitoring observations of the ultraluminous X-ray source
NGC 5204 X-1}

\author[T.\,P. Roberts et al.]
{T.\,P. Roberts$^1$\thanks{E-mail: tro@star.le.ac.uk},
R.\,E. Kilgard$^{1,2}$, R.\,S. Warwick$^1$, M.\,R. Goad$^1$ \&
M.\,J. Ward$^{3}$ \\ $^1$ X-ray and Observational Astronomy Group,
Dept. of Physics \& Astronomy, University of Leicester, University
Road, Leicester, LE1 7RH\\ $^2$ Harvard-Smithsonian Center for
Astrophysics, Cambridge, MA 02138, USA
\\ $^3$ Dept. of Physics, University of Durham, South Road, Durham DH1
3LE}
\date{}

\def\ro{{\it ROSAT~\/}}
\def\asca{{\it ASCA~\/}}
\def\ein{{\it Einstein~\/}}
\def\xmm{{\it XMM-Newton~\/}}

\def\chan{{\it Chandra~\/}}
\def\rxte{{\it RXTE~\/}}

\def\ergcms{{\rm ~erg~cm^{-2}~s^{-1}}}

\def\ergsec{{\rm ~erg~s^{-1}}}

\def\atpcm{{\rm ~atom~cm^{-2}}}
\def\ctsec{{\rm ~count~s^{-1}}}

\def\H0{{\rm ~km~s^{-1}~Mpc^{-1}}}

\def\kmsec{{\rm ~km~s^{-1}}}
\def\cmsq{{\rm ~cm^{-2}}}

\def\la{\mathrel{\hbox{\rlap{\hbox{\lower4pt\hbox{$\sim$}}}{\raise2pt\hbox{$<$}}}}}
\def\ga{\mathrel{\hbox{\rlap{\hbox{\lower4pt\hbox{$\sim$}}}{\raise2pt\hbox{$>$}}}}}

\def\d25{D$_{25}$}
\def\nh{{$N_{H}$}}

\def\.25{0.25 keV\thinspace}
\def\lx{$L_{\rm X}$}

\begin{document}

\date{In prep.}

\pagerange{\pageref{firstpage}--\pageref{lastpage}}
\pubyear{2006}

\maketitle

\label{firstpage}

\begin{abstract}
We report the results of a 2-month campaign conducted with the \chan
X-ray observatory to monitor the ultraluminous X-ray source (ULX) NGC
5204 X-1.  This was composed of a 50-ks observation, followed by ten
5-ks follow-ups spaced initially at $\sim 3$, then at $\sim 10$ day
intervals.  The ULX flux is seen to vary by factors $\sim 5$ on
timescales of a few days, but no strong variability is seen on
timescales shorter than an hour.  There is no evidence for a periodic
signal in the X-ray data.  An examination of the X-ray colour
variations over the period of the campaign shows the ULX emission
consistently becomes spectrally harder as its flux increases.  The
X-ray spectrum from the 50-ks observation can be fitted by a number of
disparate spectral models, all of which describe a smooth continuum
with, unusually for a ULX, a broad emission feature evident at 0.96
keV.  The spectral variations, both within the 50-ks observation and
over the course of the whole campaign, can then be explained solely by
variations in the continuum component.  In the context of an
optically-thick corona model (as found in other recent results for
ULXs) the spectral variations can be explained by the heating of the
corona as the luminosity of the ULX increases, consistent with the
behaviour of at least one Galactic black hole system in the
strongly-Comptonised very high state.  We find no new evidence
supporting the presence of an intermediate-mass black hole in this
ULX.
\end{abstract}


\section{Introduction}

Although ultraluminous X-ray sources (ULXs) were detected and studied
by \ein (e.g. Fabbiano 1989), \ro (e.g. Roberts \& Warwick 2000;
Colbert \& Ptak 2002; Liu \& Bregman 2005) and \asca (e.g. Makishima
et al. 2000), the \chan and \xmm observatories have provided us with
the clearest view to date of these extraordinary X-ray sources (see
e.g. Miller \& Colbert 2004 for a review of the early results from
these missions).  However, despite the advances afforded by these
missions over their predecessors, a concise determination of the
underlying nature of ULXs has remained stubbornly elusive.

One certainty is that the high observed luminosities of ULXs (\lx $~>
10^{39} \ergsec$) cannot be explained by stellar-mass black holes
radiating isotropically below their Eddington limit.  As this
luminosity limit is directly proportional to the mass of the accreting
object, the simplest solution is to turn to a new class of larger
black holes, that we now refer to as ``intermediate-mass'' black holes
(IMBHs; Colbert \& Mushotzky 1999), with typical masses in the $10^2 -
10^4$ M$_{\odot}$ range.  Such objects are certainly enticing as they
would fill in a gap in the known mass range of black holes, that
currently falls between $\sim 20$ and $\sim 10^5$ M$_{\odot}$
(i.e. between the stellar-mass objects observed in our own Galaxy, and
the supermassive black holes present in many galactic nuclei).
However, there is not a single piece of evidence that currently
establishes beyond doubt that any individual ULX contains an IMBH.  A
combination of factors - most notably its extreme luminosity - make
M82 X-1 the best current candidate (e.g. Portegies-Zwart et al. 2004).
Evidence that other bright ULXs contain IMBHs is provided largely by
the detection of a soft excess in their X-ray spectrum, that can be
fitted with an accretion disc spectrum with an inner disc temperature of
$\sim 0.1 - 0.3$ keV, suggestive of the presence of an IMBH (Miller et
al. 2003; Miller, Fabian \& Miller 2004).


However, this last spectral interpretation has recently been
challenged by new spectral data and modelling that questions the
necessity for IMBHs to power even the most luminous ULXs
(e.g. Stobbart, Roberts \& Wilms 2006; hereafter SRW06).  This
complements the circumstantial evidence from multi-wavelength data -
in particular the over-abundance of ULXs found in galaxies hosting
very active star formation - that argues against IMBHs constituting a
large fraction of the ULX population (King 2004).  Additionally, the
break in the X-ray luminosity function of ULXs at $\sim 2 \times
10^{40} \ergsec$ (Grimm, Gilfanov \& Sunyaev 2003) is difficult to
reconcile with black holes larger than $\sim 100$ M$_{\odot}$
contributing anything but a small minority of the ULX population.
These arguments therefore support a demography in which the majority
of the ULX population are a type of stellar-mass ($\sim 10$
M$_{\odot}$, though plausibly up to a few times larger) black hole
X-ray binary that can match or even exceed its Eddington limit.
Models for achieving super-Eddington luminosities from these systems
include the ``slim'' disc model (e.g. Watarai, Mizuno \& Mineshige
2001), radiation-presure dominated discs (Begelman 2002),
mildly-anisotropic radiation patterns from sources emitting at or
below the Eddington limit (King et al. 2001) and relativistic beaming
(K{\"o}rding, Falcke \& Markoff 2002).


Whilst concerted efforts have been made to understand the nature of
ULXs via X-ray spectroscopy and multi-wavelength follow-up,
variability characteristics (particularly in harness with
spectroscopy) have not yet been as widely utilised.  Yet, as {\it
RXTE\/} has so spectacularly demonstrated, the key to understanding
accreting sources is to study changes in their behaviour over a wide
variety of timescales (cf. McClintock \& Remillard 2006 for a review
of the properties of Galactic black hole binaries).  In fact,
variability measurements over many different timescales could provide
key measurements for understanding the nature of ULXs.

For example, the detection of a 3:2 ratio twin-peak high-frequency
quasi-periodic oscillation (QPO) at $\sim$ 1 Hz frequency from a ULX
would imply the presence of an IMBH (Abromowicz et al. 2004; though we
note that scaling arguments based on detections in Galactic black
holes show that even a mission with the anticipated capabilities of
{\it XEUS\/} is likely to struggle to detect such a signature).  The
detection of low-frequency QPOs may be less revealing, though several
authors have estimated the mass of the black hole in M82 X-1 based on
the frequency of a QPO detection, amongst other factors (e.g. Fiorito
\& Titarchuk 2004; Dewangen, Titarchuk \& Griffiths 2006; Mucciarelli
et al. 2006).  A simpler argument based on this QPO, which is the
best-detected example of this phenomenon in a ULX to date, is that
its presence argues against geometric beaming of the X-ray flux in M82
X-1, implying that an IMBH is likely to be present (Strohmayer \&
Mushotzky 2003; though see Kaaret, Simet \& Lang 2006b for caveats).

A second possible diagnostic for ULXs from short-timescale variability
measurements is the shape of the fluctuation Power Spectral Density
(PSD).  In particular, the characteristic frequency of breaks in the
PSD slope can be used to infer masses based on the assumption of a
direct scaling of properties between Galactic black holes and AGNs
(see e.g. Uttley et al. 2002).  Cropper et al. (2004) used this
technique to support the case that NGC 4559 X7 is a $> 1000$
M$_{\odot}$ IMBH (though they discuss alternatives).  In contrast,
Soria et al. (2004) examined the PSD of NGC 5408 X-1 during a flaring
phase and from the break frequency and slopes either side of the break
suggest a mass $\sim 100$ M$_{\odot}$.  Finally, Goad et al. (2006)
used the lack of variability power in a deep, high signal-to-noise
\xmm observation of Ho II X-1 to suggest an upper limit on its black
hole mass of $100$ M$_{\odot}$.  However, the calculation of PSDs is
not always practical for ULXs given that as a class they do not tend
to show much short-timescale variability.  For example, Swartz et
al. (2004) note short-timescale variability is detected in only 5 - 15
per cent of a large number of \chan archival observations of ULXs, and
Feng \& Kaaret (2005) only find 3 out of 28 \xmm ULX observations in
their sample show detectable noise power in a PSD analysis.

The short-timescale variability that is present in the minority of
ULXs might offer further clues as to their nature.  In particular, the
detection of periodicity could provide a view of the orbital
characteristics of the binary system underlying the ULX.  Perhaps the
most dramatic form this periodicity could take is eclipsing of the
X-ray emission by the secondary star; interestingly, this could also
be an indicator of the black hole mass, with stellar-mass black holes
far more likely to show eclipses than IMBHs (Pooley \& Rappaport
2005).  However, known eclipsing ULXs remain rare -- David et
al. (2005) report a candidate in NGC 3379 with a 8-10 hr period, and a
second candidate with a 7.5-hr period has long been known in the
Circinus galaxy (Bauer et al. 2001) (though this latter object may
instead be a foreground AM Her system; Weisskopf et al. 2004).  Other
reports of periods mainly rely on apparent periodic flux maxima in the
light curves of the ULXs (for example Liu et al. 2002; Kaaret, Simet
\& Lang 2006a), though these detections are often based on very
limited data (coverage of 2 - 4 times the suggested period only).  The
ambiguities in interpreting such data is highlighted by the case of
M74 X-1 (CXOU J013651.1+154547), which was observed to undergo strong,
relatively regular flaring behaviour during \chan and \xmm
observations in 2001/2002.  Liu et al. (2005) interpret this as
quasi-periodic behaviour, and from this and the X-ray spectrum suggest
a mass of $\sim (2 - 20) \times 10^{3}$ M$_{\odot}$, whereas Krauss et
al. (2005) note that the behaviour resembles Galactic microquasars and
suggest that the flaring could be related to a relativisitic jet
aligned with our line-of-sight.

The general lack of short-timescale variability exhibited by ULXs is
to some degree extended to longer timescales by their persistent
brightness.  Indeed, many bright ULXs first observed by \ein and
subsequently re-observed by the following missions have remained
active for more than 20 years with only modest flux variations between
epochs (factors $< 4$ over $\sim 20$ years, e.g. Roberts et al. 2004).
Such persistence is reminiscent of Galactic high-mass X-ray binaries
with black hole primaries.  Interestingly, Kalogera et al. (2004)
predict that accretion from massive stars onto both stellar-mass black
holes and IMBHs will appear reasonably persistent, with the key
diagnostic being that IMBH systems will show transient behaviour on
timescales of tens of years\footnote{In Kalogera et al., the
behavioural differences are anticipated to be between IMBHs of mass $>
50$ M$_{\odot}$ and less massive black holes, for accretion from
fairly massive ($> 7$ M$_{\odot}$) young secondary stars.}  No such
complete turn-off has been seen so far from a famous ULX.  Indeed, the
majority of well-studied transient ULXs to date are likely to be
similar in nature to known Galactic sources (Ghosh et al. 2006a; Mukai
et al. 2005; Bauer \& Pietsch 2005), though the spectrally hard, very
luminous ULX close to the centre of NGC 3628 stands out as an IMBH
candidate that has shown possible transient behaviour (Strickland et
al. 2001).

A final temporal diagnostic is the spectral behaviour of ULXs,
particularly by comparison to the behaviour of Galactic black hole
X-ray binaries.  \asca observations of the ULXs in IC 342 identified
spectra that apparently transited between low/hard (power-law
dominated) and high/soft (accretion disc dominated) states in
observations separated by 7 years (Kubota et al. 2001), which is
similar to the most common behaviour in Galactic systems.  However,
many subsequent studies have highlighted ULXs that appear to show a
contrary behaviour, in that they spectrally harden as their flux
increases (e.g. Fabbiano et al. 2003; Dewangan et al. 2004).  Fabbiano
et al. (2003) note that this may indicate that ULXs are in a
particularly high accretion rate state, though Jenkins et al. (2004)
caution that the apparent difference may in part be due to the choice
of X-ray colour bands used to define the hardness ratios.

Here, we present the results of a 2-month \chan monitoring campaign on
NGC 5204 X-1, that probes the variability diagnostics of this ULX on
timescales from seconds to weeks.  NGC 5204 X-1 is one of the
brightest nearby ULXs with historical observed fluxes in the range
$\sim 0.7 - 2 \times 10^{-12} \ergcms$ (0.5 -- 8 keV), converting to
luminosities of the order $2 - 6 \times 10^{39} \ergsec$ for a
distance of $d = 4.8$ Mpc (Roberts et al. 2004).  It is most notable
for possessing stellar optical counterparts identified from its
accurate \chan position (Roberts et al. 2001; Goad et al. 2002), one
of which was subsequently identified as a B0 Ib supergiant and
suggested to be the actual counterpart due to UV spectral
peculiarities consistent with some Galactic X-ray binaries (Liu,
Bregman \& Seitzer 2004).  Initial short \chan observations were
hampered by low-level pile-up, but did reveal a steep power-law X-ray
spectrum (Roberts et al. 2001, 2004).  Subsequent \xmm observations
then revealed a spectral soft excess that might be interpreted as
evidence of an IMBH accretion disc, though other spectral models not
requiring the presence of an IMBH also fitted well to the data
(Roberts et al. 2005; SRW06).  In this paper we return to the issue of
the nature of this ULX, as constrained by our X-ray monitoring
campaign.

\section{The monitoring campaign}

Previous observations of NGC 5204 X-1 have indicated that its flux
varies by factors of $\sim 3$ over a baseline of years, but that
little or no measureable variability was seen over timescales of
minutes to a few hours (e.g. Roberts et al. 2004).  Hence this
monitoring campaign was designed to provide an indication of whether
this ULX varies on timescales intermediate to those already probed,
i.e. timescales of days to weeks.  The \chan X-ray observatory is the
best choice for such a programme primarily due to its scheduling
flexibility, combined with the spectral imaging capabilities of
ACIS-S.  The variability was investigated by a sequence of 11
observations spanning a period of two months (Table~\ref{obsdet}).  An
initial moderately-deep (50-ks) observation began the sequence,
providing information on variability over the course of a $\sim 14$
hour interval.  This was followed by a sequence of five short (5-ks)
follow-up observations spaced at $\sim 3$ day intervals, to
investigate variability over the course of days, then a further five
(5-ks) observations spaced at $\sim 10$ day intervals to look at the
variability on timescales of weeks.  We summarise the observation
details in Table~\ref{obsdet}, showing the exact start times and
live-time corrected exposures.  We also show an estimate of the mean
\chan ACIS-S count rate from NGC 5204 X-1 for each observation, which
clearly demonstrates that the source flux is indeed varying over the
day-to-week timescales examined.

\begin{table}
\caption{Details of the \chan observations.}
\begin{tabular}{lcccc}\hline
Sequence number & Date and start time	& Exposure	& Count rate \\
 & (yyyy-mm-dd@hh:mm:ss) & (s)	& (count s$^{-1}$) \\\hline 
600307	& 2003-08-06@05:58:27	& 46227	& 0.20 \\
600308	& 2003-08-09@09:36:20	& 4737	& 0.11 \\
600309	& 2003-08-11@19:10:34	& 4500	& 0.13 \\
600310	& 2003-08-14@16:35:34	& 4562	& 0.47 \\
600311	& 2003-08-17@07:49:47	& 4635	& 0.45 \\
600312	& 2003-08-19@23:28:06	& 5149	& 0.11 \\
600313	& 2003-08-27@17:18:30	& 5143	& 0.11 \\
600314	& 2003-09-05@16:11:45	& 4841	& 0.28 \\
600315	& 2003-09-14@11:41:35	& 4877	& 0.46 \\
600316	& 2003-09-23@09:33:01	& 5207	& 0.25 \\
600317	& 2003-10-03@23:31:14	& 4896	& 0.41 \\
\hline
\end{tabular}
\label{obsdet}
\end{table}

A second aim of this programme was to characterise any spectral
variability over the timescales probed.  However, as Roberts et
al. (2004) demonstrated, even employing the ACIS-S ${{1}\over{8}}$
subarray mode in an attempt to mitigate pile-up was not fully
successful for previous \chan ACIS-S observations of NGC 5204 X-1.
When the ULX was located on-axis, and was at its higher flux state ($>
0.4 \ctsec$), detectable pile-up effects were still evident in its
X-ray spectrum.  We therefore configured each observation such that
NGC 5204 X-1 was located a distance of 4 arcminutes off-axis,
displaced along the ACIS-S3 chip such that it remained in the middle
of the spatial region covered by the ${{1}\over{8}}$ subarray (i.e. an
offset of $\Delta$Y $= - 4$ arcmin in detector coordinates).  The
degradation of the \chan point spread function at $4$ arcminutes
off-axis results in a 50\% encircled energy radius of $\sim 1.5$
arcseconds, which compares to $\sim 0.5$ arcseconds on-axis.  This is
sufficient to ensure that any remaining pile-up is mitigated to
manageable levels ($< 4\%$).

The resulting data were reduced using standard tools within the
\textsc{ciao} 3.0 software suite, and version 2.23 of the Chandra
Calibration Database.  Pulse invariant spectra were extracted for each
observation using the tool \textsc{dmextract} and response matrices
and ancillary response functions (ARFs) were computed.  Due to the
build-up of the contaminant layer on the ACIS optical blocking filter,
the ARFs were corrected using Alexey Vikhlinin's tool
\textsc{corrarf}, now incorporated as a standard part of the
\textsc{ciao} spectral extraction.

In order to assess the validity of these products in light of
calibration improvements made publicly available since our initial
reduction, we reprocessed data from the 50-ks observation using
\textsc{ciao} 3.3 and the calibration database (\textsc{caldb}) 3.2.1,
and extracted new spectra.  Minor differences in the response
functions were found, primarily below 0.5 keV, with the functions
being equivalent above 1 keV.  Given that the impact to data in the
0.5-1 keV range was minimal, we elected not to reprocess the data
presented herein.


\section{Temporal variability}

\begin{figure}
\centering
\includegraphics[width=5cm,angle=270]{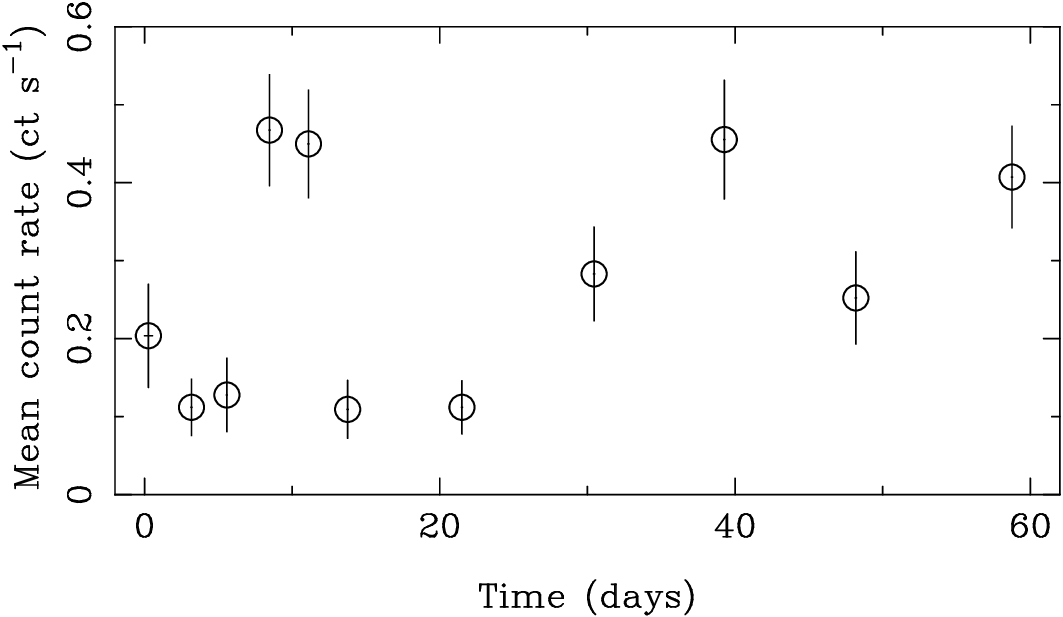}
\caption{Full light curve for the monitoring campaign.  We show the
mean 0.5 -- 8 keV band count rate per observation, and use the
standard deviation on this count rate to represent the intrinsic
variability per epoch.}
\label{fulllc}
\end{figure}

\begin{figure*}
\centering
\includegraphics[width=17.5cm,angle=0]{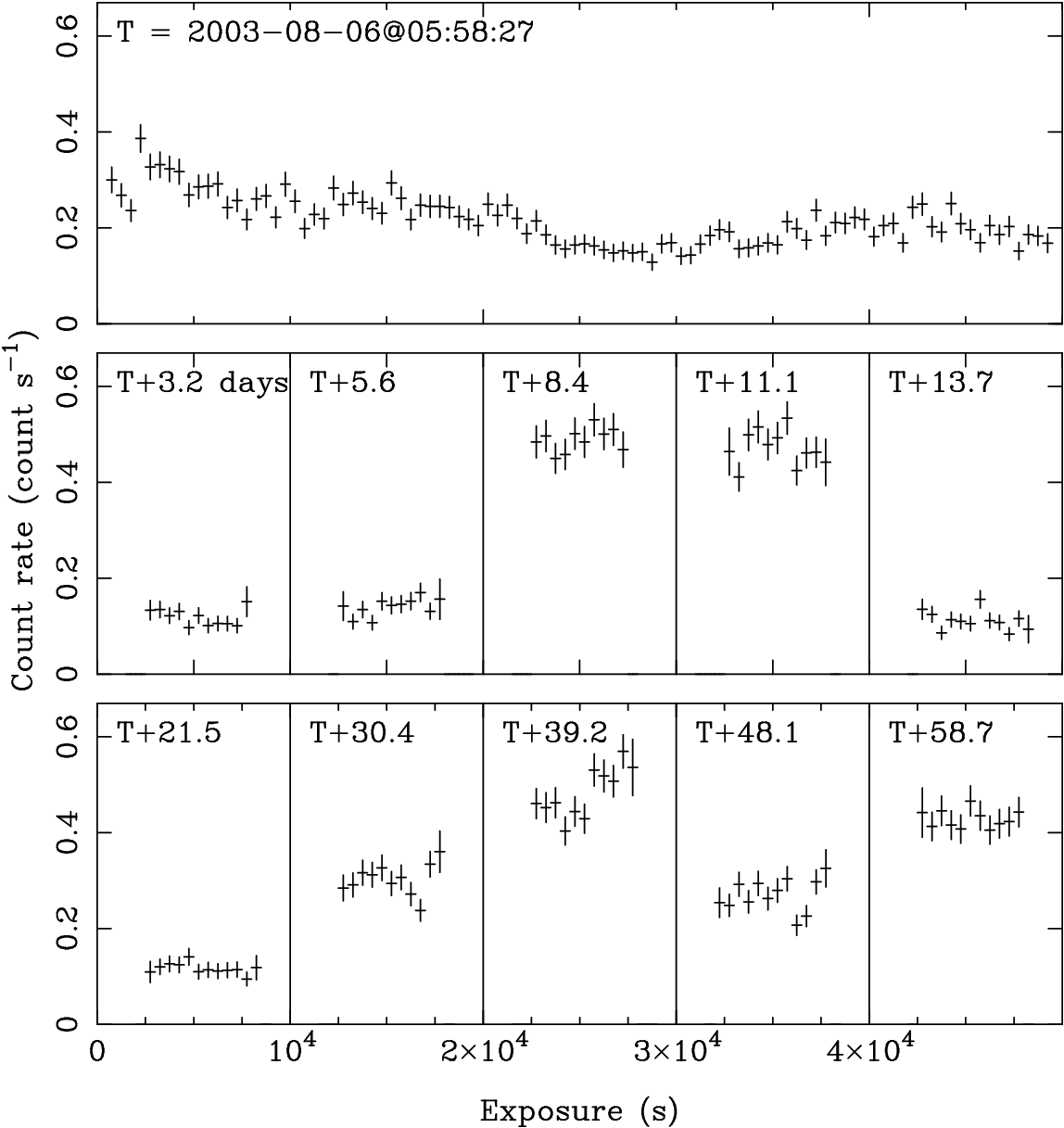}
\caption{The full light curve for the monitoring programme of NGC
5204 X-1, binned to 500 s resolution.}
\label{indvlcs}
\end{figure*}

We present the 2-month monitoring light curve of NGC 5204 X-1 in
Figure~\ref{fulllc}.  The mean observed count rate of the ULX clearly
varies by as much as factors of $\sim 5$ (in the 0.5 - 8 keV
\chan bandpass) between the observations, and the variations are
statistically highly significant (a $\chi^2$ fit against the
hypothesis of a constant count rate, set to the mean count rate for
the eleven observations, gives a statistic of 95.5 for 10 degrees of
freedom).  Most interestingly, the source appears to have upper and
lower thresholds that are not exceeded (albeit within the very limited
sampling of this programme), confining the flux measurements to
between $\sim 0.1$ and $0.5 \ctsec$.  In fact, these limits do not
appear to have been exceeded throughout the historical record of
observations of this source (cf. Roberts et al. 2004, Figure 6(f),
where the extreme flux points are both from \chan ACIS-S data, and
correspond to count rates of $\sim 0.41$ and $\sim 0.16 \ctsec$).

The light curve, though sparsely sampled, does not betray any obvious
periodic signature.  This is particularly interesting for NGC 5204 X-1
where a positive identification of the secondary star in this system
as a B0 Ib supergiant, and the assumption of Roche lobe overflow,
leads to firm predictions of the orbital period of between $\sim 200 -
300$ hours for a wide range of primary black hole masses ($3 - 1000$
M$_{\odot}$; Liu et al. 2004).  Our $\sim 59$ day baseline covers
between 5 and 7 cycles of these possible orbital periods.  We have
searched for possible hidden periodic signals by deriving a
Lomb-Scargle periodogram from the data.  A visual inspection of the
periodogram suggested possible periodic signals of $\sim 10$ and 16
days (i.e $\sim 240$ and 384 hours).  We further investigated these
signals by fitting a model composed of a sinusoidal function plus a
constant component to the lightcurve, and did indeed find a best-fit
to the data at $\sim 10$ days.  However this fit produced a $\chi^2$
statistic of 27.4 for 7 degrees of freedom (dof), which has a null
hypothesis probability of only $3 \times 10^{-4}$, hence there is very
little chance of this being a true periodic signal.  The fit for a
16-day period was even worse ($\chi^2$/dof = 39.5/7).  We conclude
that there is little or no evidence for periodicity in the X-ray
emission from this dataset.

In Figure~\ref{indvlcs} we show the eleven individual lightcurves at
500-s resolution, plotted on a convenient scale for direct comparison.
An inspection by eye reveals that as one looks at shorter timescales
the amplitude of variability in this source is apparently less.  That
is to say, during the 50-ks observation clear trends are visible,
starting with a diminution in count rate from $\sim 0.3 \ctsec$ in the
first 5 ks to $\sim 0.15 \ctsec$ after $\sim 30$ ks, followed by a
slight brightening to $\sim 0.2 \ctsec$ for the remainder of the
observation.  However, trends and/or large amplitude variations are
less visible within the shorter follow-up observations.  We performed
standard statistical tests to check for significant intrinsic
variability (i.e. a $\chi^2$ test against the hypothesis of a constant
count rate for data binned to 100 or 200 second resolution, and a
Kolmogorov-Smirnov test at the full 0.74104 s temporal resolution of
the ACIS-S3 ${{1}\over{8}}$ subarray), and found that the only
observation showing highly significant variability was indeed the
50-ks look (sequence number 600307, $\gg 10\sigma$ significance in
both tests).  Only one other observation showed variability at $>
3\sigma$ significance, namely 600315 (at T + 39.2 days) that was
significant at the $\sim 4\sigma$ level according to the K-S test
(though, in contrast, the $\chi^2$ test showed a significance $<
2\sigma$ for this observation).  There was no strong evidence of any
short-term variability intrinsic to the data in any other observation.

\begin{figure}
\centering
\includegraphics[width=6cm,angle=270]{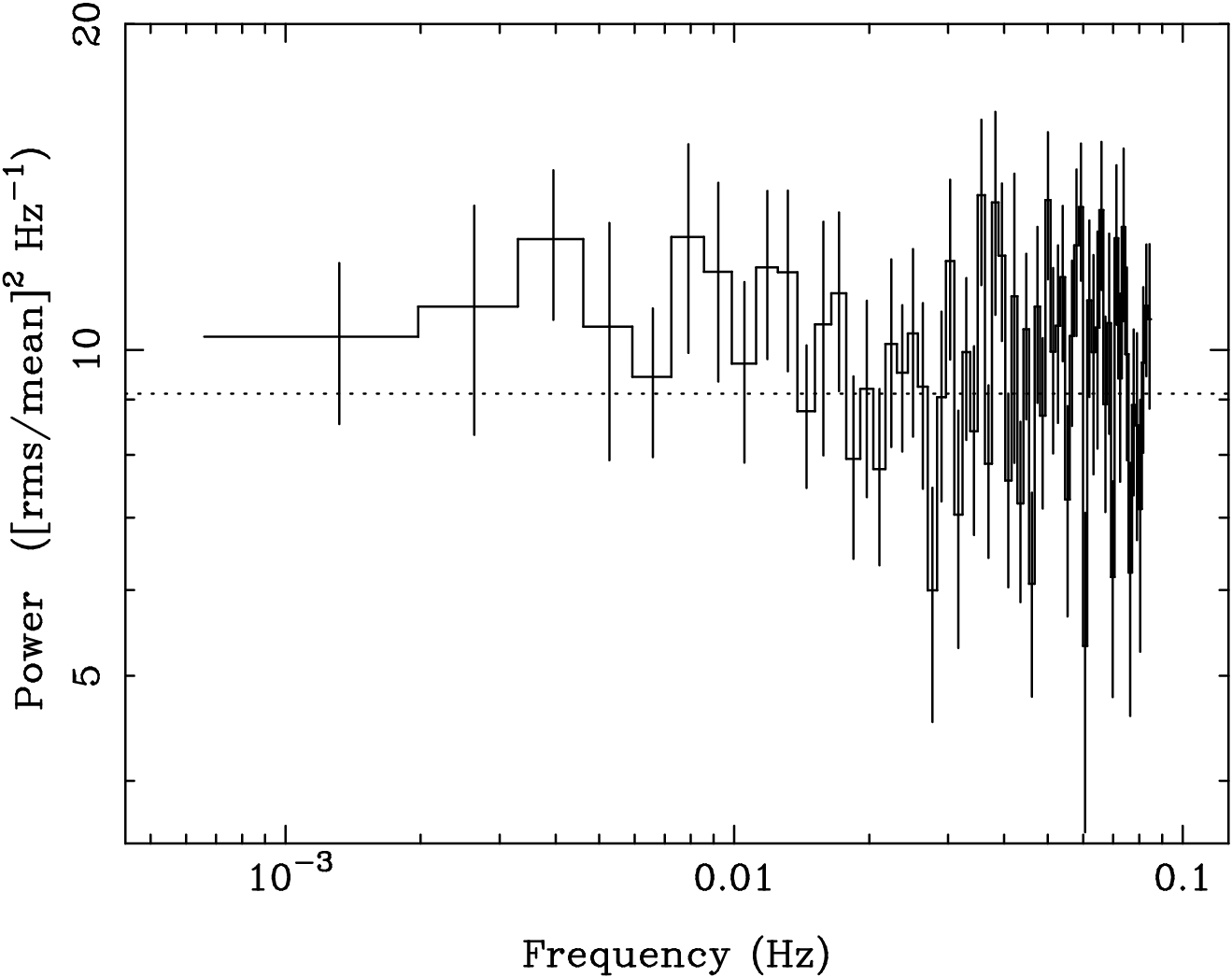}
\caption{The power spectral density of the 50-ks exposure.  It was
calculated by dividing the light curve into 64 segments of 128 points,
and the FFT of each segment was computed and averaged to produce the
plot.  Errors were computed in the standard fashion from the
dispersion of the 64 data points within each frequency bin. The PSD
was normalised using the (rms/mean)$^2$ convention defined by van der
Klis (1997).  Using this normalisation the expected Poisson noise
level (shown with the dotted line) is $2/\mu$ (where $\mu$ is the mean
count rate).}
\label{psd}
\end{figure}

The increase in intrinsic variability over longer timescales is
typical of a source showing a red noise dominated fluctuation power
spectral density (PSD).  We have checked this by calculating the PSD
from the longest (50-ks) observation, which we show in
Figure~\ref{psd}.  The PSD is clearly poisson-noise dominated above
0.02 Hz, though some excess power may be evident below this frequency
(i.e. on timescales $> 100$ s), consistent with a red noise source
spectrum.  However, the data for this potential red noise component
are so poor that we are unable to constrain any possible measurement
of the power-law slope characterising this component at the 90\%
level.  We note that as the variability increases to factors up to
$\sim 5$ over time scales of $\sim 3$ days the putative red noise
spectrum should continue down to, and become more apparent at,
frequencies longer than those covered in the 50-ks observation.

\begin{figure*}
\centering
\includegraphics[width=6cm,angle=270]{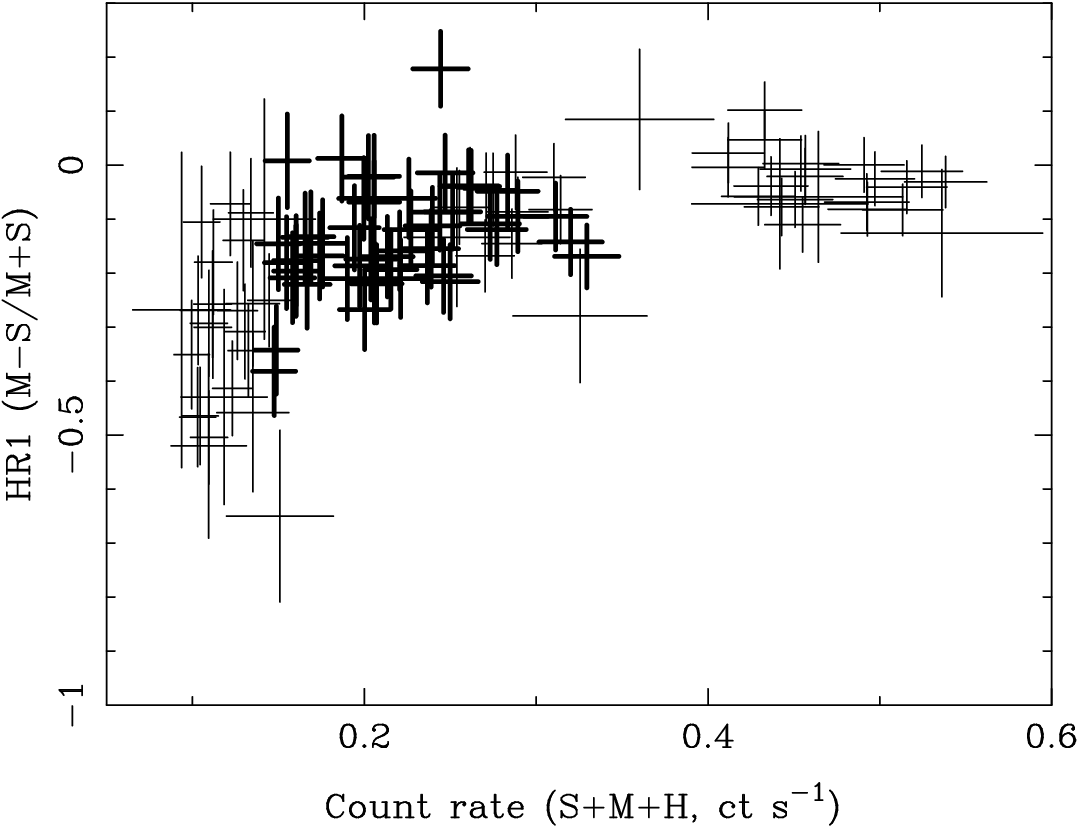}\hspace*{1cm}
\includegraphics[width=6cm,angle=270]{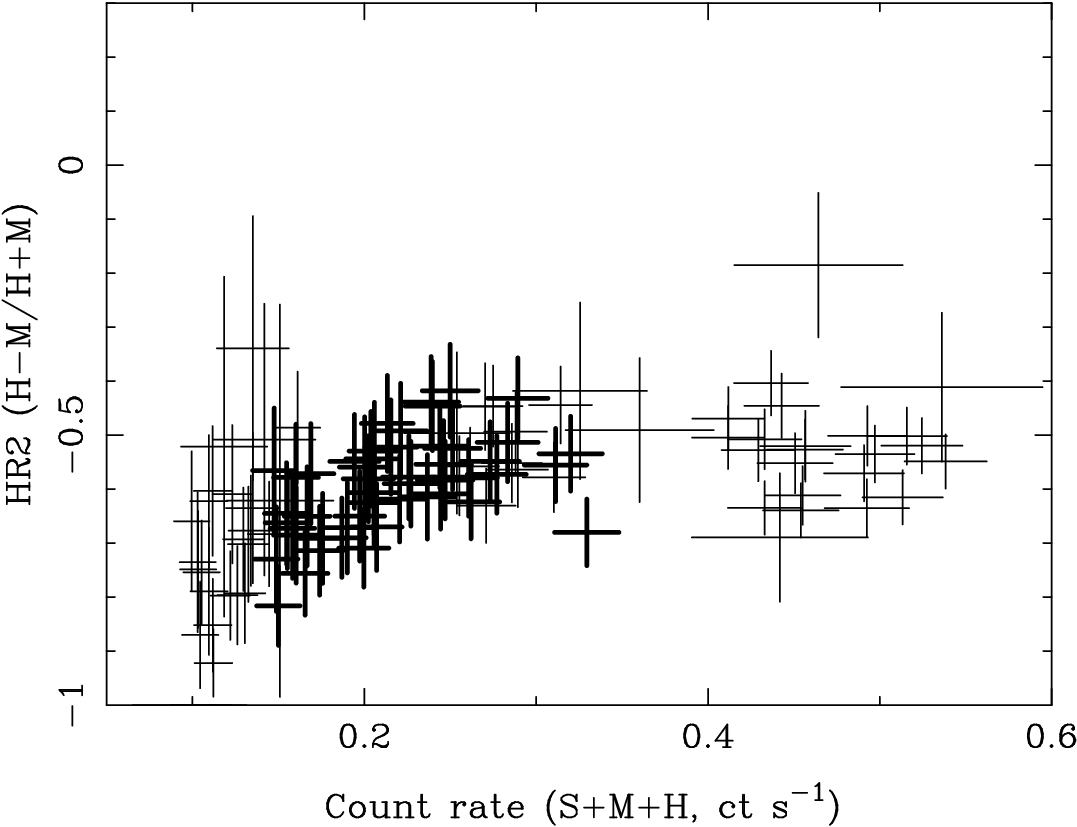}
\caption{Variations in the spectral hardness ratios of NGC 5204 X-1
with overall count rate.  The hardness ratios are calculated in 1 ks
intervals, as per the method discussed in the text.  We plot data from
the deep (50 ks) observations as thick crosses, with the data from the
ten shorter follow-up observations as thin crosses.}
\label{hrs}
\end{figure*}



\begin{table}
\caption{Mean hardness ratio values within flux-limited ranges.}
\begin{tabular}{lccc}\hline
Range	& Count rates (ct s$^{-1}$)	& HR1	& HR2 \\\hline
Faint	& $< 0.2$	& $-0.23 \pm 0.01$	& $-0.71 \pm 0.02$ \\
Medium	& $0.2 - 0.4$	& $-0.11 \pm 0.01$	& $-0.55 \pm 0.01$ \\
Bright	& $> 0.4$	& $-0.03 \pm 0.01$	& $-0.54 \pm 0.01$ \\
\hline
\end{tabular}
\label{hrvals}
\end{table}

We have also investigated whether the broad spectral form of the ULX
alters during our monitoring campaign by examining spectral hardness
ratios calculated from the data.  We binned the data into 1-ks
intervals and segregated it into three energy bands, a 0.3 -- 1 keV
soft ($S$) band, a 1 -- 2 keV medium ($M$) band, and a 2 -- 8 keV hard
($H$) band.  We then calculated our hardness ratios as HR1 $=$
$(M-S)/(M+S)$ and HR2 = $(H-M)/(H+M)$, with errors on the hardness
ratios calculated as per Ciliegi et al. (1997).

An initial examination of the energy-segregated lightcurves suggested
that there may be spectral variations as the total flux of the source
changes.  We investigated this further by comparing the hardness
ratios to the total count rate of NGC 5204 X-1, which we show here in
Fig.~\ref{hrs}.  There does indeed appear to be a trend over the
course of the campaign, with the ULX appearing to be relatively soft
in its lowest flux states, and to harden significantly as its flux
increases.  This appears to be the case for both hardness ratios.
Remarkably this behaviour appears contrary to the classic low/hard to
high/soft state transitions commonly seen in Galactic black hole
binaries though, as we noted earlier, similar behaviour has been
observed before in other ULXs.  We quantify this relationship in
Table~\ref{hrvals}, where we split the data into three broad bins
based on the total count rate, and calculate a weighted mean of the
hardness ratio values within that count rate range.  In the softer of
the two ratios, HR1, we see that the average source spectrum hardens
over all three bins.  However, the average HR2 value is consistent
between the two brighter count rate bins, and only significantly
softer in the faintest bin.  We investigate the nature of these
spectral variations, and their dependence on the brightness of NGC
5204 X-1, in the next section.

\section{Spectral variability}

Spectral data were extracted in an elliptical aperture, with major and
minor axes of $\sim 7.8$ and $\sim 4.3$ arcseconds respectively,
designed to encompass the full \chan point spread function for X-ray
photons from NGC 5204 X-1 at $\sim 4$ arcminutes off-axis.  Background
data were extracted in an elliptical annulus, centered on the ULX and
with inner and outer axial dimensions of 1 and 4 times the source
extraction ellipse axes respectively.  The spectral data were grouped
to a minimum of 20 counts per bin, to ensure $\chi^2$ statistics are
valid.  Spectral fitting was then performed over the 0.5 -- 10 keV
range\footnote{Data below 0.5 keV were not considered in this analysis
due to calibration uncertainties in this regime.  In addition,
although we quote 10 keV as the upper limit, in practise we have only
one data bin above 6 keV.}, using \textsc{xspec} v.11.3.0.  The errors
quoted throughout the analysis are 90\% errors for one interesting
parameter unless otherwise stated.

\subsection{The X-ray spectrum from the 50-ks observation}

We began our spectral analysis by examining the full spectral data set
for NGC 5204 X-1 from the initial 50-ks observation.  In this and the
subsequent analysis we fitted two absorption columns: a Galactic
foreground absorber set to the line-of-sight in the direction of NGC
5204 ($1.5 \times 10^{20} \atpcm$; Stark et al. 1992), and an
additional absorption column presumably located along the
line-of-sight in NGC 5204, or in the local vicinity of the ULX.  In
both cases we used the interstellar abundances and absorption
cross-sections tabulated by Wilms, Allen \& McCray (2000) (the
\textsc{tbabs} model in \textsc{xspec}).

\begin{figure}
\centering
\includegraphics[width=5.5cm,angle=270]{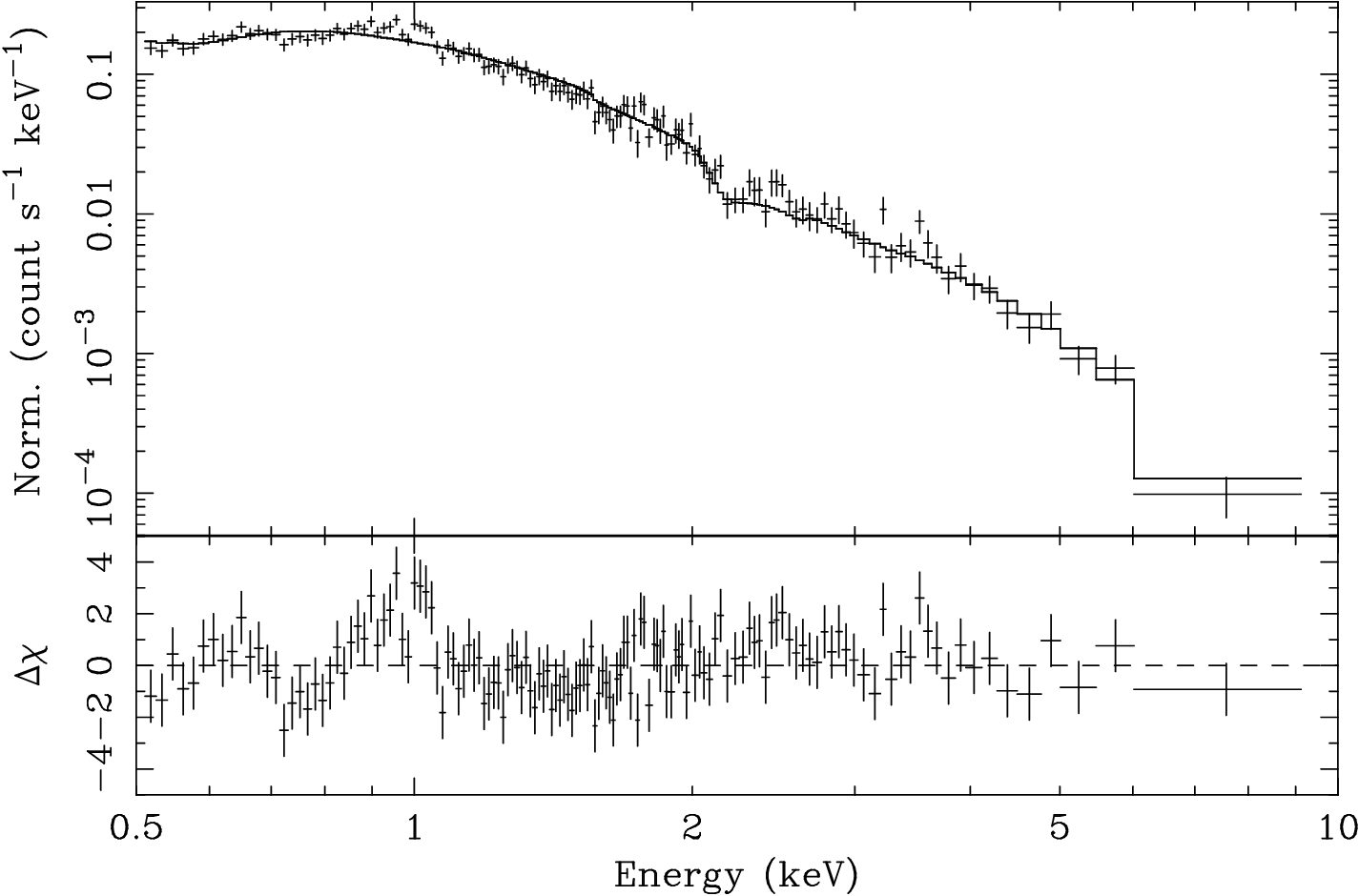}
\caption{Power-law continuum fit to the spectrum of the 50 ks
observation.  The upper panel shows the data and best fitting
power-law model (see text for details), and the lower panel shows the
$\Delta\chi$ residuals to this model.}
\label{plfit}
\end{figure}

Initial fits to the data were attempted using absorbed,
single-component spectral models.  However, despite trying a wide
range of simple models (e.g. blackbody; thermal bremmstrahlung;
multi-colour disc blackbody; \textsc{mekal} optically-thin thermal
plasma) no good fit to the data was found.  The best-fitting simple
model was an absorbed power-law continuum, as found in previous
observations of NGC 5204 X-1 (Roberts et al. 2001, 2004, 2005), in
this case with \nh $=$ $(1.3 \pm 0.2) \times 10^{21} \cmsq$ and
$\Gamma = 2.83 \pm 0.08$.  However this was still rejected at high
significance ($\chi^2$/dof of 212.3/141, i.e. a rejection probability
of $P_{rej} =$ $99.99\%).$\footnote{We consider spectral fits to be
formally acceptable if their rejection probability is $< 95\%$.}  We
show the power-law fit and residuals in Fig.~\ref{plfit}.  As with
previous observations, the overall continuum shape of this ULX is
soft.  However, this spectrum is remarkable in that it shows a large
residual feature at $\sim 0.8 - 1$ keV, with further possible
residuals evident at both lower and higher energies.

As we cannot achieve good fits using single component spectral models,
we turned to two-component models.  Specifically, we tried the
empirical models discussed at length in Roberts et al. (2005),
utilised in fitting the \xmm spectrum of this source.  These were the
standard empirical description of a black hole X-ray binary spectrum,
a soft multicolour disc blackbody plus a harder power-law continuum,
and the non-standard inversion of this spectrum where the power-law
continuum dominates at soft energies.  We note that the \xmm data
showed ambiguity between these two models, i.e. it did not indicate a
strong preference for one or the other.  We again find that our data
does not strongly distinguish between these two models for NGC 5204
X-1, with best fitting parameters of inner accretion disc temperature
$kT_{in} \sim 0.18$ keV and power-law photon index $\Gamma \sim 2.7$
for the standard model fit, and $kT_{in} \sim 1.2$ keV, $\Gamma \sim
3.9$ for the non-standard fit.  However, unlike the \xmm observation,
neither model provides a good description of the data in this case,
with both rejected at 99.96\% significance.

Instead, statistically-acceptable fits were found using two-component
models in which the soft component was modelled by a \textsc{mekal}
thermal plasma.  Good fits (rejection probability $P_{rej} < 80\%$)
were found using either a power-law continuum or a multi-colour disc
blackbody as the hard component.  We detail these fits as Models (1)
and (2) in Table~\ref{specfits}.  When a power-law continuum describes
the harder emission the \textsc{mekal} temperature is $\sim 0.96$ keV,
and its abundance is approaching solar, whereas when the multi-colour
disc blackbody models the hard component we require a cooler ($\sim
0.84$ keV), low metallicity \textsc{mekal}.

\begin{table*}
\centering
\caption{Spectral modelling of the 50 ks \chan observation of NGC 5204 X-1.}
\begin{tabular}{llccccc}\hline
\multicolumn{2}{l}{Model component $^a$}	& \multicolumn{5}{c}{Spectral models $^a$}\\
	&	& (1)	& (2) 	& (3) 	& (4) 	& (5) \\
	&	& \textsc{mekal} + \textsc{po}	& \textsc{mekal} + \textsc{diskbb}	& \textsc{bb} + \textsc{po} + \textsc{gauss}	& \textsc{bb} + \textsc{diskbb} + \textsc{gauss}	& \textsc{comptt} + \textsc{gauss} \\\hline
\textsc{tbabs}	& \nh $~^b$	& $< 1.2$	& $< 0.05$	& $0.98^{+0.68}_{-0.58}$	& $< 0.19$	& $< 0.45$ \\
\textsc{mekal}	& $kT ~^c$	& $0.96 \pm 0.09$	& $0.84^{+0.05}_{-0.06}$	& - 	& - 	& - \\
		& $Z ~^d$	& $0.62^{+2.92}_{-0.52}$	& $0.04 \pm 0.01$	&  - 	& - 	& - \\
\textsc{po}	& $\Gamma ~^e$	& $2.55^{+0.12}_{-0.40}$	&  -	& $2.92^{+0.23}_{-0.17}$ 	& - 	& - \\
\textsc{diskbb}	& $kT_{in} ~^f$	& -	& $1.43^{+0.25}_{-0.18}$	& -	& $0.92 \pm 0.06$	& - \\
		& $A_{dbb} ~^g$	& - 	& $(4.4^{+3.7}_{-2.2}) \times 10^{-3}$	& -	& $(4.6^{+1.6}_{-1.2}) \times 10^{-2}$	& - \\
\textsc{bb}	& $kT ~^h$	&  -	& - 	& $0.67^{+0.12}_{-0.13}$ 	& $0.17 \pm 0.01$ 	& - \\
\textsc{gauss}	& $E ~^i$	& - 	& - 	& $0.96 \pm 0.02$ 	& $0.97 \pm 0.02$ 	& $0.96 \pm 0.02$ \\
		& $\sigma ~^j$	& - 	& - 	& $0.08^{+0.02}_{-0.03}$	& $0.06 \pm 0.03$ 	& $0.07 \pm 0.03$ \\
		& EqW$ ~^k$	& - 	& - 	& $85^{+22}_{-40}$	& $51^{+23}_{-17}$	& $75^{+33}_{-21}$ \\	
\textsc{comptt}	& $T_0 ~^l$	& -  	& - 	& -	& - 	& $0.11^{+0.01}_{-0.04}$ \\
		& $kT_e ~^m$	& - 	& - 	& -	& - 	& $1.41^{+0.37}_{-0.35}$ \\
		& $\tau ~^n$	& - 	& - 	& -	& - 	& $6.7^{+1.6}_{-1.3}$ \\
$\chi^2/$dof	& & $150.6/138$	& $151.6/138$	& $140.5/136$	& $145.7/136$	& $139.1/136$ \\\hline	
\end{tabular}
\begin{tabular*}{\textwidth}{p{17cm}}
Notes: $^a$ In \textsc{xspec} syntax.  All models are subject to an
additional, fixed Galactic column of $1.5 \times 10^{20}$ cm$^{-2}$.
$^b$ Absorbing column external to our own Galaxy, in units of
$10^{21}$ cm$^{-2}$.  $^c$ Plasma temperature (keV).  $^d$ Plasma
metallicity (solar units).  $^e$ Power-law photon index.  $^f$
Inner-disc temperature (keV).  $^g$ Model normalisation.  $^h$
Black-body temperature (keV).  $^i$ Line centroid energy (keV).  $^j$
Line width (keV).  $^k$ Equivalent width of line (eV).  $^l$ Seed
photon temperature for corona (keV).  $^m$ Comptonising corona
electron temperature (keV).  $^n$ Optical depth of corona.
\end{tabular*}
\label{specfits}
\end{table*}

In order to investigate whether any specific features were driving the
{\sc mekal} fits to the soft part of the spectrum, we proceeded to fit
the spectral data with a combination of smooth continuum models and
emission lines.  As we were able to fit the hard component with either
a power-law or a multi-colour disc blackbody in Models (1) \& (2), we
therefore attempted two continuum fits, one with each as the hard
component.  We chose to use a classic blackbody as the soft component.
This results in two empirical models with possibly quite different
physical origins, i.e. an accretion disc plus hot corona model
(blackbody + power-law continuum), and an accretion disc plus a soft
excess model (in which the soft excess is modelled by the blackbody).
The use of both models resulted in good fits to the data above 1 keV,
but left large residuals at lower energies as expected.  In both cases
we were able to fit these residuals, and hence provide an acceptable
fit to the full dataset, using a single broad gaussian line at an
energy of $\sim 0.96$ keV (cf. Models (3) \& (4) of
Table~\ref{specfits}).  However, neither underlying continuum model
appears to support the presence of a cool ($\sim 0.1 - 0.3$ keV)
accretion disc in this ULX, that one would expect if an IMBH were
present.

There has been some recent speculation that ULXs may form
preferentially in low-metallicity environments, as these potentially
allow the formation of large black holes (cf. Soria et al. 2005 and
references therein).  Evidence in support of this has recently emerged
in a deep \xmm observation of the Ho II ULX (Goad et al. 2006), where
an apparent excess of counts above 0.5 keV in the RGS spectrum of this
ULX was best explained by reducing the metallicity of the absorbing
material to $\sim 0.6$ of the solar values.  If we allow the
metallicity of the second of our absorption components to vary (by
using the \textsc{tbvarabs} model, with all abundances constrained to
the same value as a first approximation) we find that a sub-solar
abundance does provide a marginal improvement to the fit for Model
(3), though as Model (4) includes little or no extra absorption above
the Galactic foreground column it is insensitive to this test.  The
actual improvement for Model (3) is $\Delta\chi^2 =5.1$ for one extra
degree of freedom, which is marginally significant (at the $\sim
2\sigma$ level) according to the F-test.  The parameterisation of
Model (3) is essentially unchanged, apart from a substantially higher
column of $4.5^{+2.6}_{-2.3} \times 10^{21} \cmsq$, and a limit of $<
0.26$ times solar abundance for the metallicity of the absorbing
medium.

\begin{figure}
\centering
\includegraphics[width=5.5cm,angle=270]{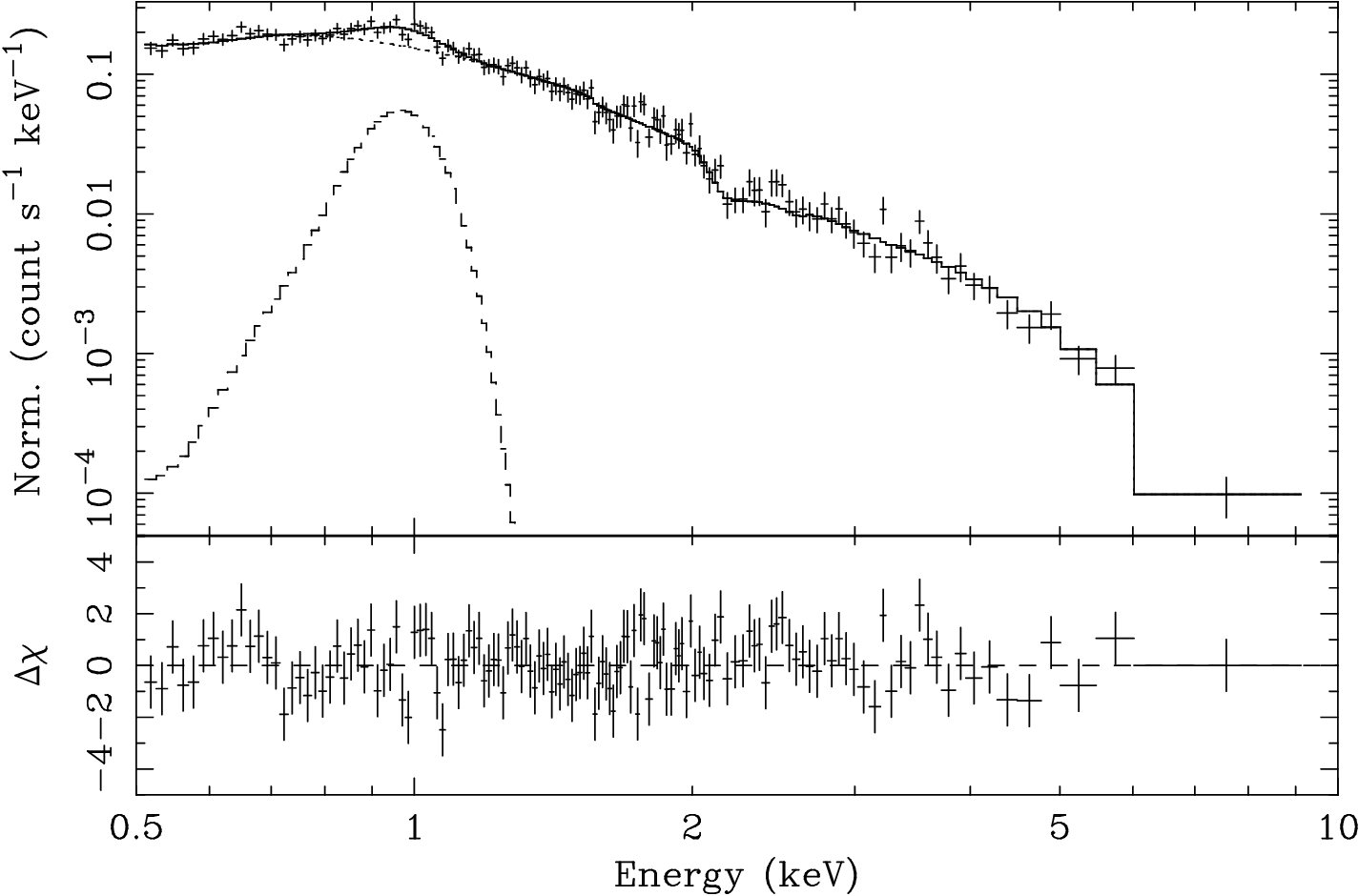}
\caption{Model (5) fitted to the spectrum of the 50 ks observation.  The
upper panel shows the data and best fitted model, with the contribution
of the gaussian line and the optically-thick corona highlighted by
dashed and dotted lines respectively (as can be seen, the corona
effectively models the full continuum other than in the 0.8 - 1.1 keV
regime).  The lower panel shows the $\Delta\chi$ residuals to this
model.}
\label{cttfit}
\end{figure}

Finally, we attempted to fit a physical accretion disc plus corona
model to the data, as one might expect to see for an accreting black
hole.  As per the work of Goad et al. (2006) on Ho II X-1, we used a
\textsc{diskpn} + \textsc{comptt} model (subject to the same
solar-abundance absorption components as the empirical models), with
the two components linked by using the temperature of the disc to
provide the input temperature of the seed photons for the corona.
This model did not provide a statistically-acceptable fit to the data,
with the best fit having $\chi^2$/dof $= 194/138$, though it did have
the interesting characteristic of being totally dominated by the
coronal component.  An inspection of the residuals to this fit showed
that the feature at $\sim 1$ keV was not adequately explained by this
model; we therefore again added a Gaussian line.  Similarly to the
empirical models, this provided a statistically-acceptable fit to the
data, which we show as Model (5) of Table~\ref{specfits}.  In fact, it
provides the statistically best fit to the data, albeit by a small
margin.\footnote{This is not simply due to allowing more degrees of
freedom to vary in the Comptonisation model than are present in
empirical models - in fact, this fit has the same number of degrees of
freedom as models (3) and (4).}  We show the data and Model (5) fit in
Fig.~\ref{cttfit}. Even after the addition of the line - which shows
very similar properties to the line derived from the empirical fits -
the contribution of the disc component is negligible ($\ll 1\%$ of the
0.5 -- 8 keV flux), so we only show the parameterisation of the
coronal component in the Table.  This component shows the same
remarkable characteristics as the corona in Ho II X-1, i.e. it is cool
($kT_e \sim 1.4$ keV), optically-thick ($\tau \sim 7$)\footnote{The
{\sc comptt} model is not fully self-consistent, and may underestimate
the actual depth of the corona.  For example, SRW06 model the Ho II
X-1 corona using the more physical {\sc eqpair} model, and find an
optical depth a factor 2 deeper than Goad et al.'s {\sc comptt}
modelling.} and seeded by photons originating in a cool ($\sim 0.1$
keV) disc.  Interestingly, this is supportive of other recent work,
most notably a sample of 13 bright ULXs observed with \xmm (SRW06),
that shows that it is plausible that the X-ray spectra of a large
proportion of ULXs are well described by an optically-thick coronal
component seeded by a cool disc.  We note that by fixing $kT_e = 50$
keV and re-fitting, one can get a statistically-acceptable solution
with an optically-thin corona ($\tau \sim 0.2, \chi^2/$dof $=
150.6/137$), though this does require the disc temperature to reduce
further to $\sim 20$ eV.  However, an F-test shows that the
improvement offered over this by the optically-thick solution is
significant at the $\sim 3\sigma$ level.  We therefore view an
optically-thick solution as far more realistic.

The spectral modelling results in a typical observed 0.5 -- 8 keV flux
of $\sim 9.8 \times 10^{-13} \ergcms$ (with only a few per cent
disagreement between the models in Table~\ref{specfits}), from which
we can derive an average luminosity of $\sim 2.7 \times 10^{39}
\ergsec$ over the duration of the observation.

\subsection{Spectral variability within the 50 ks observation}

\begin{figure}
\centering
\includegraphics[width=4.5cm,angle=270]{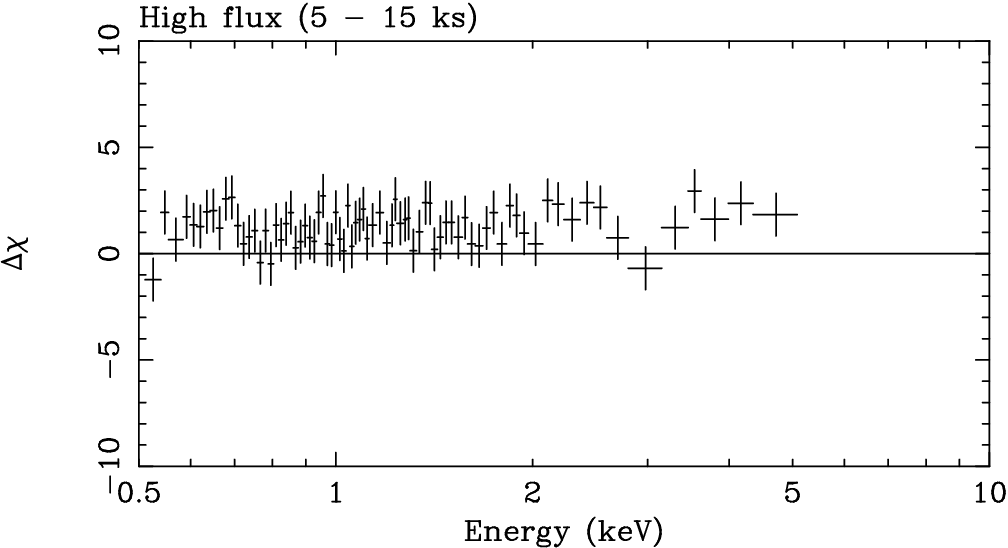}\vspace*{0.2cm}
\includegraphics[width=4.5cm,angle=270]{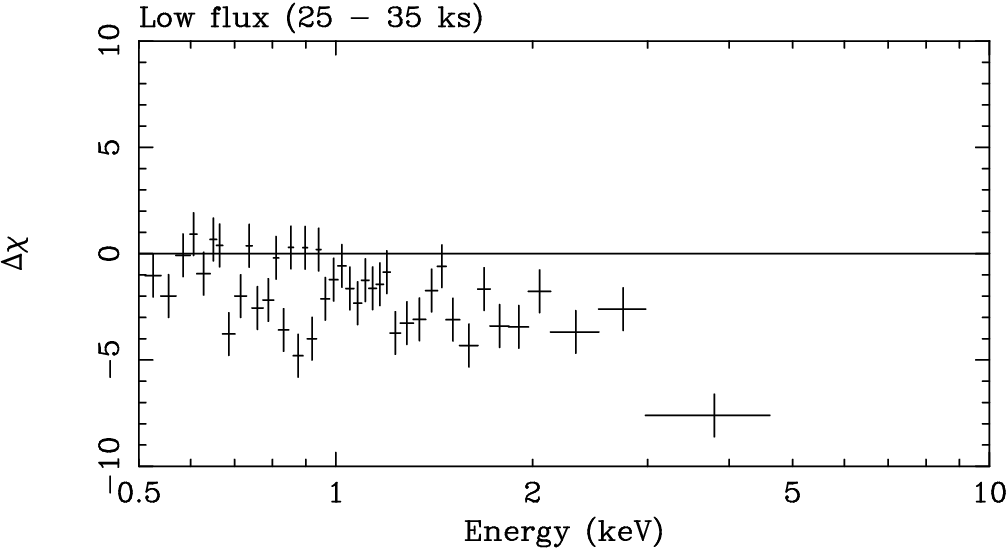}\vspace*{0.2cm}
\includegraphics[width=4.5cm,angle=270]{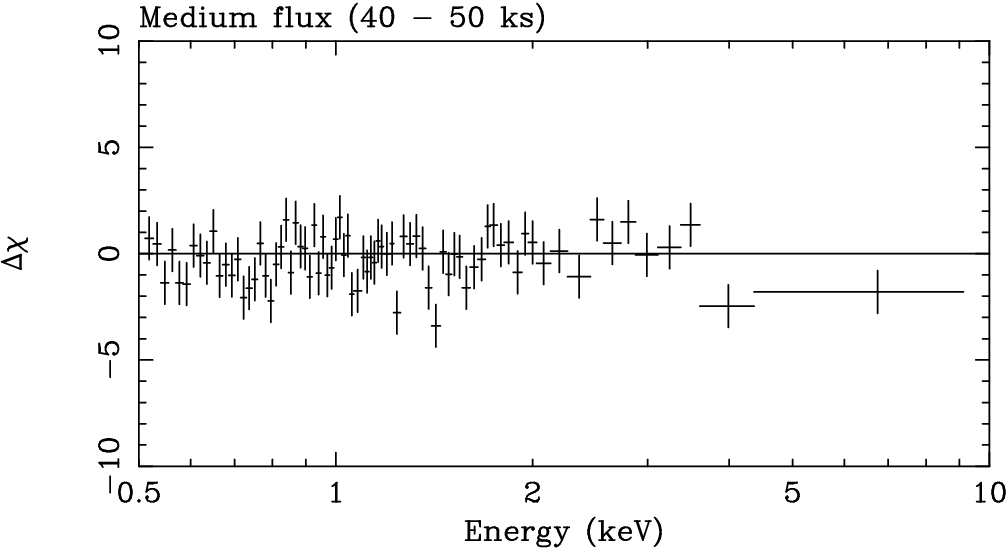}
\caption{The spectral variability of NGC 5204 X-1 during the 50 ks
observation.  We use the three segments described in the text, and
plot the $\Delta\chi$ residuals against the Model (5) fit, i.e. the
best-fitting description of the spectrum of the whole observation.
The three panels are shown on the same scale for direct comparison.}
\label{varrats}
\end{figure}

As Fig.~\ref{hrs} shows, the factor $\sim 2$ variation in flux during
the observation is accompanied by some spectral variation.  We have
investigated whether we can quantify this variation through spectral
fitting in the following manner.  We first selected and extracted
spectral data from three 10-ks segments within the 50-ks observation,
these being the 5 -- 15, 25 -- 35 and 40 -- 50 ks windows.  Roughly
speaking, these correspond to relatively stable high ($> 0.2 \ctsec$),
low ($< 0.2 \ctsec$) and medium ($\sim 0.2 \ctsec$) flux states during
the observation (cf. Fig.~\ref{indvlcs}; we refer to each segment by
these respective titles hereafter).  The relative differences in these
spectra are demonstrated in Fig.~\ref{varrats}, where we plot the
$\Delta\chi$ residuals for each time segment spectrum against the best
fitting model to the whole observation (Model (5), i.e. the
optically-thick coronal model).

In order to study these spectral differences, we loaded all three
datasets into \textsc{xspec} simultaneously.  The varying components
of each spectral model were identified by first freezing all the model
components, and then thawing the parameters one-by-one, untying their
values across the three datasets, and re-fitting.  Those parameters
offering the largest reduction in the overall $\chi^2$ value for the
simultaneous fit were judged to be the most important in explaining
the spectral variation of the ULX.

As Fig.~\ref{varrats} shows, a comparison of the spectral segments to
the model best-fitting the whole spectrum demonstrates the presence of
considerable residuals.  Indeed, when loaded into \textsc{xspec} this
model proved a very poor simultaneous fit to the three data segments,
with a $\chi^2$/dof value of $572/200$.  The majority of spectral
variability appears to originate above 1 keV, particularly in the case
of the ``low'' segment spectrum, where considerable additional
curvature is present.  We find that this spectral variability can be
very simply and adequately modelled by a variation in the temperature
of the coronal component, which provides a statistically-acceptable
fit to the data ($\chi^2$/dof $= 226.7/197$, i.e. $P_{rej} \approx
93\%$).  A striking feature of this fit is that it appears that the
temperature of the coronal component falls and rises in line with the
flux of the ULX, that is to say it is at its coolest when the flux is
lowest, and vice versa (see Table~\ref{varyspec}).  A slight
additional improvement to the fit is offered by thawing and untying
the optical depth of the corona, though as it offers a mere
$\Delta\chi^2 = 6.9$ improvement for three additional degrees of
freedom it is not statistically robust.  However, we still tabulate
the best-fitted values of $\tau$ in Table~\ref{varyspec} (alongside
the best-fitted coronal temperatures and fluxes for each segment) as
we take the view that changes in the optical depth are at least a
plausible occurrence as the coronal temperature varies.  Furthermore,
we note that freeing either the temperature of the seed photons, or
the normalisation of the coronal model, does not provide any
improvement to the fit.  Finally, in this model there is no necessity
for the flux, central line energy or line width of the Gaussian
component at 1 keV to vary at all during the observation.

\begin{table}
\caption{Varying spectral parameters within the 50-ks observation.}
\begin{tabular}{lcccc}\hline
Segment (ks)	& $kT_{\rm e} ~^a$	& $\tau ~^b$ 	& $f_{\rm X} ~^c$	& $L_{\rm X} ~^d$ \\\hline
5 -- 15		& $1.74 \pm 0.09$	& $6.2 \pm 0.4$		& $13.2^{+0.3}_{-0.4}$	& $3.7 \pm 0.1$ \\
25 -- 35	& $1.10 \pm 0.09$	& $6.9^{+0.8}_{-0.7}$ 	& $6.2 \pm 0.2$	& $1.8 \pm 0.1$  \\
40 -- 50	& $1.29 \pm 0.07$	& $7.1 \pm 0.5$ 	& $9.0^{+0.3}_{-0.2}$	& $2.6 \pm 0.1$  \\\hline
\end{tabular}
\begin{tabular}{l}
Notes: $^a$ Temperature of corona in keV.  $^b$ Optical depth of
corona.\\ $^c$ Observed 0.5 -- 8 keV flux, in units of $\times
10^{-13} \ergcms$.\\ $^d$ Intrinsic 0.5 -- 8 keV luminosity, in units
of $\times 10^{39} \ergsec$.
\end{tabular}
\label{varyspec}
\end{table}

\subsection{Spectral variability throughout the programme}

\begin{figure}
\centering
\includegraphics[width=4.5cm,angle=270]{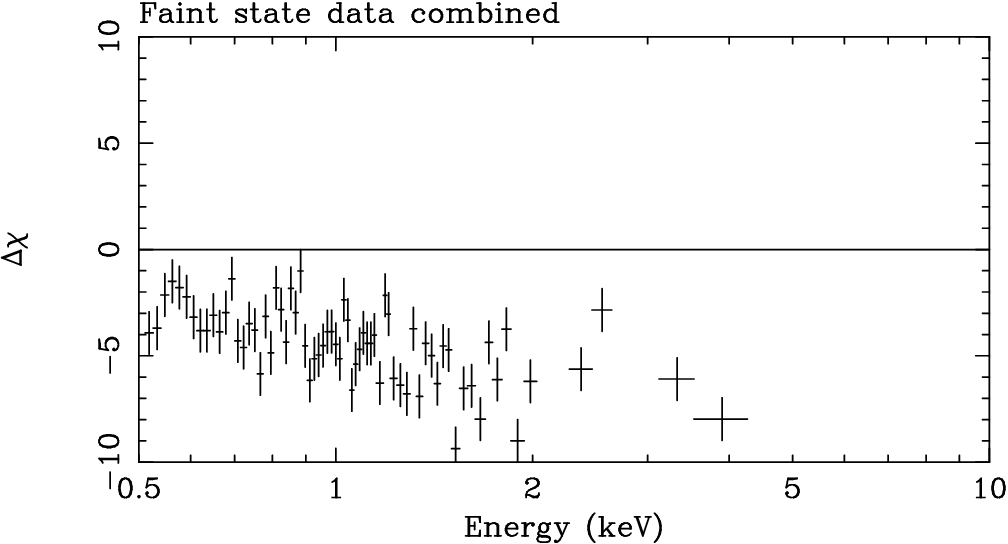}\vspace*{0.2cm}
\includegraphics[width=4.5cm,angle=270]{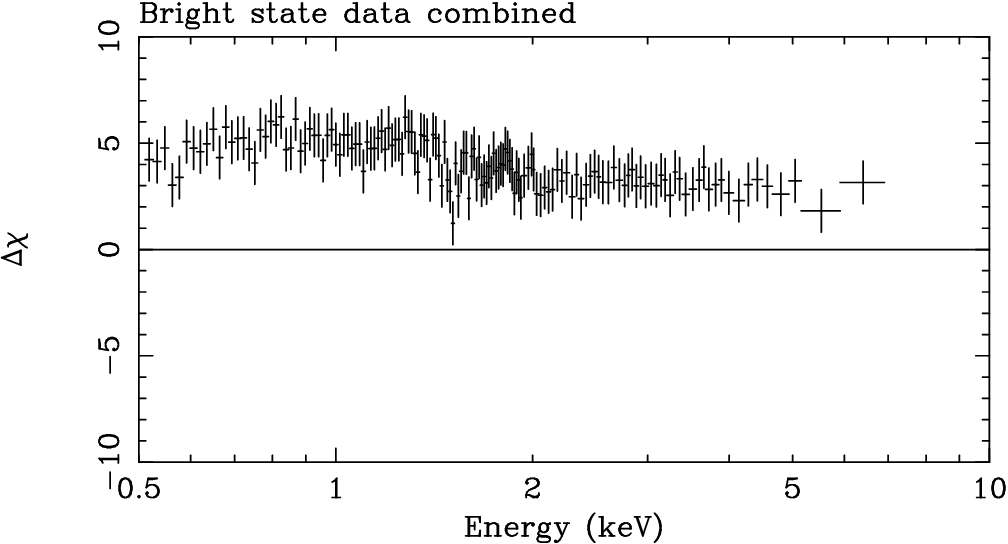}
\caption{The spectral variability of NGC 5204 X-1 from the follow-up
observations.  As per Fig.~\ref{varrats}, we plot the $\Delta\chi$
residuals against the Model (5) fit.  The panels are shown on the same
scale as the previous Fig. for direct comparison, though we note that
this excludes two data points from the top panel in the 2 -- 3 keV
regime, that lie at $\Delta\chi \sim -15$.}
\label{varrats2}
\end{figure}

As Fig.~\ref{hrs} indicates, the spectrum of NGC 5204 X-1 seems to
vary in a consistent manner as the count rate varies across the entire
dataset, including the series of follow-up observations.  We therefore
hypothesize that the observed X-ray variability during the entire
monitoring campaign can be explained by the optically-thick corona
model (i.e. Model (5)), as proved adequate for the intra-observation
variability during the 50-ks observation.  Unfortunately, the data
from the 5-ks snapshots are generally of insufficient quality to
constrain the fits (particularly when the count rate is low).  We
therefore co-added spectra to produce reasonable quality data to work
with.  Specifically, we co-added spectra from observation sequence
numbers 600308, 600309, 600312 and 600313, to provide a ``faint''
state spectrum, as all these observations had average count rates in
the 0.11 -- 0.13 $\ctsec$ regime.  Similarly, we combined spectra from
observation sequence numbers 600310, 600311, 600315 and 600317
(i.e. where average count rates were between 0.41 -- 0.47 $\ctsec$),
to provide a ``bright'' state spectrum.  Again, we began by comparing
the extracted spectra to the best fitted spectrum from the deep
observation (Model (5)), and note that both co-added spectra showed
considerable changes relative to this spectrum.  We again demonstrate
this by plotting the $\Delta\chi$ residuals of these datasets compared
to Model (5) in Figure~\ref{varrats2}.

We find that the faint state spectrum is very adequately explained by
the optically-thick corona model, with $\chi^2$/dof $= 67.8/69$.  The
majority of the variability can be modelled as a change in the coronal
temperature, with additional marginally significant improvements (at
the $\sim 95\%$ level according to the F-test) offered by changes in
the optical depth of the corona and, for the first time, changes in
the temperature of the photons seeding the corona (i.e. the accretion
disc temperature).  Interestingly, the 90\% errors on the parameters
do allow an optically-thin, hot corona solution in this case, but
again only if the accretion disc temperature reduces further to $\sim
20$ eV.  We note that the presence of the Gaussian line is required in
this fit - indeed, setting its normalisation to zero degrades the
goodness of the fit significantly ($\Delta\chi^2 = 50$).

The bright state cannot be as readily explained using Model (5).
Thawing each of the coronal temperature, its optical depth and seed
photon temperature in turn provides a significant improvement to the
fit each time, though the resultant fit has $\chi^2$/dof $= 195/146$,
i.e. it is rejected at the 99.6\% confidence level.  However, an
inspection of the residuals reveals that just two isolated bins (at
$\sim 1.5$ and $\sim 5.5$ keV) disagree strongly with the model,
providing $\Delta\chi^2 \approx 18$ each to the fit.  If they are
removed, the fit becomes tenable ($\chi^2$/dof $= 159/146$,
i.e. $P_{rej} < 80\%$).  We therefore conclude that, assuming these
two bins do not represent real spectral features\footnote{The deficit
present in the $\sim 5.5$ keV bin can be modelled by an edge feature,
at an energy of $5.1 \pm 0.2$ keV, with a depth of $0.9^{+0.6}_{-0.4}$
keV.  This produces an improvement of $\Delta\chi^2 = 20.4$ for two
extra degrees of freedom, and results in a fit that is only marginally
rejected ($P_{rej} = 95.8\%$).  However, an edge at this energy may be
astrophysically implausible - the nearest edge is neutral Ti ($\sim 5$
keV), or if it were to be an edge from neutral Fe it would require a
redshift of $\sim 0.3$.  No such improvement is possible for the 1.5
keV bin, which is unlikely to be a real feature as its width is far
narrower than the ACIS-S resolution at that energy.  There are no
known ACIS-S calibration issues at this energy, and an inspection of
the individual contributory spectra indicates that this feature
originates in the juxtaposition of less significant residual data
points from two out of four spectra, indicating this feature may
simply be an extreme statistical fluctuation.}, the underlying
spectral form is still adequately modelled by the optically-thick
corona model.

We compare the spectral parameters from the bright and faint co-added
spectra with the parameters from the 50-ks observation (derived whilst
the Gaussian parameters and absorption remained frozen, as is the case
for the co-added spectra) in Table~\ref{varyspec2}.  In this
comparison we see apparent differences in the temperature of the seed
photons (i.e. the underlying accretion disc) between the epochs when
the ULX is at different mean fluxes.  As these flux changes only occur
on timescales of days, it is interesting to note that this may
indicate that the accretion disc temperature changes are occuring on
the same timescale.  It is certainly true that in the one instance we
are sensitive to shorter timescale variations in the accretion disc
temperature, i.e. in the 50-ks observation, we do not see such an
effect.  We note again that the temperatures - in this case of both
the seed photons and the corona - appear to vary in line with the
overall flux of the spectrum, with the system being hotter when it is
brighter, though the optical depth appears to vary independently in
this comparison.

\begin{table}
\caption{Varying accretion disc plus corona spectral parameters over
the campaign, and for an earlier \xmm observation.}
\begin{tabular}{lccccc}\hline
Obs.	& $T_0 ~^a$	& $kT_{\rm e} ~^b$	& $\tau ~^c$ 	& $f_{\rm X} ~^d$	& $L_{\rm X} ~^e$ \\\hline
50-ks	& $0.11 \pm 0.01$	& $1.41^{+0.13}_{-0.11}$	& $6.7 \pm 0.4$		& $9.8^{+0.1}_{-0.2}$	& $2.8^{+0.0}_{-0.1}$ \\
Faint	& $0.09^{+0.02}_{-0.08}$& $1.22^{+67}_{-0.28}$		& $6.2^{+1.2}_{-5.2}$ 	& $4.6^{+1.3}_{-0.8}$	& $1.3^{+0.4}_{-0.2}$ \\
Bright	& $0.14 \pm 0.01$	& $2.29^{+0.14}_{-0.12}$	& $5.1 \pm 0.2$ 	& $23^{+11}_{-3}$	& $6.4^{+3.0}_{-0.8}$ \\
	& \\
{\it XMM\/}	& $0.11^{+0.03}_{-0.02}$	& $1.99^{+0.38}_{-0.24}$	& $7.5^{+0.7}_{-0.8}$	& $14^{+2}_{-6}$	& $3.9^{+0.6}_{-1.7}$ \\\hline
\end{tabular}
\begin{tabular}{l}
Notes: $^a$ Temperature of photons seeding the corona (keV) $^b$
Temperature\\ of corona in keV.  $^c$ Optical depth of corona.  $^d$
Observed 0.5 -- 8 keV flux,\\ in units of $\times 10^{-13} \ergcms$.
$^e$ Intrinsic 0.5 -- 8 keV luminosity,\\ in units of $\times 10^{39}
\ergsec$.
\end{tabular}
\label{varyspec2}
\end{table}

Whilst this and the previous section have demonstrated that the
spectral variability of NGC 5204 X-1 can readily be explained through
changes in the dominant coronal component of Model (5), this was not
the only spectral model capable of providing an acceptable fit to the
full data from the 50-ks observation.  By repeating the above analysis
for the other spectral models in Table~\ref{specfits} we find that
Models (1), (3) and (4) can all also adequately describe the spectral
variations (we exclude Model (2) for reasons explained below in
Section 5.2).  We tabulate the variable parameters for these models in
Table~\ref{varyempspec}.  In each case, the component responsible for
the $\sim 1$ keV feature - a \textsc{mekal} in Model (1), or Gaussian
line in Models (3) \& (4) - remains invariant, whilst the other
parameters are required to change to provide an acceptable fit to the
data.\footnote{In fitting the models, we allowed the power-law
normalisations for Models (1) and (3), and the normalisation for the
blackbody in Model (4), to vary.  A variation in the normalisations
for the components plausibly describing an accretion disc - the
blackbody in Model (3), and the multi-colour disc blackbody in Model
(4) - was not required, consistent with changes in the temperature of
the inner-edge of the accretion disc.}.  In each case, the fitted
parameter values again reflect the pattern of spectral hardening with
increased luminosity.

\begin{table*}
\caption{Varying empirical model parameters over the campaign.}
\begin{tabular}{lccccccc}\hline
	& \multicolumn{7}{c}{Model \& parameters $^a$} \\
Obs.	& (1)	& & \multicolumn{2}{c}{(3)}	& & \multicolumn{2}{c}{(4)} \\
	& $\Gamma$	& & $kT$	& $\Gamma$	& & $kT$	& $kT_{\rm in}$ \\\hline
50-ks	& $2.55 \pm 0.04$ 	& & $0.67^{+0.08}_{-0.07}$& $2.93 \pm 0.05$& & $0.17 \pm 0.01$	& $0.92 \pm 0.01$ \\
Faint	& $3.10^{+0.14}_{-0.13}$& & $0.48^{+0.08}_{-0.07}$& $3.98^{+0.17}_{-0.16}$& & $0.16 \pm 0.01$	& $0.73 \pm 0.01$ \\
Bright	& $2.35 \pm 0.04$	& & $0.60^{+0.13}_{-0.10}$& $2.56 \pm 0.04$& & $0.20 \pm 0.01$	& $1.14 \pm 0.01$ \\
	& \\
{\it XMM\/}	& $2.08 \pm 0.02$	& & $1.22^{+0.09}_{-0.08}$	& $2.67^{+0.04}_{-0.06}$	& & $0.18 \pm 0.01$	& $1.54^{+0.08}_{-0.07}$\\\hline
\end{tabular}
 \\
Notes: $^a$ Models and parameters as per Table~\ref{specfits}.  
\label{varyempspec}
\end{table*}

\subsection{A comparison with \xmm data}

As a final metric of the longer term spectral variations of NGC 5204
X-1 we have re-analysed data from an \xmm observation of NGC 5204 X-1
taken 8 months before our monitoring programme commenced.  This data
set (Observation ID 0142270101) was taken on 2003 January 6, and
provides 15.4/18.5 ks of live-time corrected exposure in the EPIC
pn/MOS instruments respectively, with data quality comparable to the
50-ks \chan observation.  (We note that there were two further
observations of NGC 5204 X-1 taken in mid-2003, but both were subject
to strong background flaring leaving little useful exposure for
analysis.)  The results of this observation have previously been
discussed by Roberts et al. (2005) and SRW06.  Here we use the \xmm
science analysis software ({\textsc{sas}) version 6.5.0 to re-extract
the data from the original observation data files (ODFs) available in
the \xmm science archive, in order to take advantage of refinements in
the calibration of the EPIC instruments.  We then created 0.3 -- 10
keV source spectra from this data, and fit them in \textsc{xspec}, in
the manner described by Roberts et al. (2005).

The data were firstly fitted with Model (5) using the same methodology
as in the previous sections, i.e. the best fitting parameters from the
50-ks \chan observation were thawed one-by-one to assess whether
variations from the parameter in question best explain the difference
between data sets.  In this case we were required not just to vary the
three physical parameters used to describe the variations in the \chan
data ($T_{\rm 0}, kT_{\rm e}$ and $\tau$), but also the normalisations
of the disc and coronal components, in order to achieve an acceptable
fit.  We show the best fit parameterisation of the \xmm spectra using
Model (5) in Table~\ref{varyspec2}.  The \xmm data appears to follow
the same flux - coronal temperature relationship as observed for the
various \chan data.  Unlike the \chan data, we were able to measure an
appreciable contribution to the spectrum from an accretion disc
component in the \xmm data, likely due to the slightly softer response
of EPIC compared to ACIS.  However, this was small, amounting to only
$\sim 2.5$ per cent of the observed flux in the 0.3 -- 10 keV band,
though as SRW06 found the disc contribution in empirical disc plus
corona models rarely amounts to more than 10 per cent of the flux in
this band for ULXs.

As in the previous section, we also fit the \xmm data with Models (1),
(3) and (4).  The results are shown in Table~\ref{varyempspec}.
Again, we required that extra normalisations were thawed and re-fit in
order to get statistically-acceptable fits, notably the normalisations
for the disc components in Models (3) and (4).  The parameterisation
of the \xmm fits emphasize that this spectrum appears different to the
subsequent \chan observations.  Overall, they betray a harder
spectrum, with both harder spectral parameters, and a general
dimunition of the flux in the softer component.  For example, the flux
in the \textsc{mekal} component in Model (1) is reduced to $\sim 20$
per cent of its value in the later \chan observations (though, where
present, the flux of the Gaussian feature is again not required to
change in order to obtain a good fit).  Hence it appears that in the
eight months between the \xmm observation and the \chan monitoring the
spectrum of NGC 5204 X-1 softened, in large part due to an increase in
flux at low energies.  Finally, we emphasize that these models are
just a sub-set of those fitting this data, and refer the reader to
Roberts et al. (2005) and SRW06 for more discussion.

\section{Discussion}

\subsection{On the temporal variability}

As this programme is amongst the first dedicated X-ray monitoring
campaigns for any ULX, it provides us with a new view of a
poorly-explored region of ULX parameter space, namely X-ray variations
over timescales of hours to weeks.  During the campaign the count rate
of NGC 5204 X-1 varied between $\sim 0.1 - 0.5$ count s$^{-1}$, with a
corresponding observed luminosity range of $\sim 1.3 - 6.3 \times
10^{39} \ergsec$.  The most obvious feature to look for in such data
is a periodicity on the timescale of days.  Indeed, using the B0 Ib
identification of the counterpart to NGC 5204 X-1 found by Liu et
al. (2004), and assuming that the star (which has $M_* = 25$
M$_{\odot}$, $R_* = 30$ R$_{\odot}$) fills its Roche lobe, Liu et
al. predict orbital periods of 200-300 hours for black holes in the
mass range 3-1000 M$_{\odot}$ underlying NGC 5204 X-1, providing us
with a testable hypothesis.

The most likely and dramatic manifestation of this periodicity would
be in the form of eclipses of the X-ray source by the supergiant
secondary star, as is discussed for ULXs in general by Pooley \&
Rappaport (2005).  These authors point out that the presence of
eclipses in ULX light curves could be an additional diagnostic of the
mass of the black hole in the ULX, with stellar-mass black hole
systems at least twice as likely to display eclipses.  Promisingly,
the fairly large amplitude (factor $\sim 5$) variability seen in NGC
5204 X-1 on timescales of days occurs between fairly consistent
``high'' ($\sim 0.5$ count s$^{-1}$) and ``low'' ($\sim 0.1$ count
s$^{-1}$) flux states, that one might perhaps expect to see as the
source goes in and out of eclipse.  However, no periodic signal to
support this interpretation was recovered from the data.  This argues
that the variability is not due to eclipsing, but may simply be due to
accretion processes producing variability on the timescale of days.
If this is true of many ULXs, then detecting X-ray eclipses on the
timescale of their orbits (which Pooley \& Rappaport note could vary
from 1-150 days, even for a very limited choice of black
hole/secondary star mass assumptions) may be far more difficult than
simply getting round the logistical problem of acquiring X-ray
monitoring data with a high observational efficiency over timescales
of days.\footnote{Though we note that true extragalactic eclipsing
sources may now be starting to be observed, for example with the
detection of a slightly sub-ULX luminosity source in NGC 4214 with an
eclipse periodicity of 3.6-hr (Ghosh et al. 2006b).}

Despite the lack of a periodic signal, short-timescale
(i.e. intra-observation) variability could still be a useful
diagnostic of the accretion state of the ULX through standard analysis
techniques, for example the calculation of the fluctuation Power
Spectral Density for the ULX.  However, such analyses are in practice
hampered by the poor photon statistics of our data - we have too few
photons for reliable PSD analyses of the short observations, and can
do no better than a few 10s of seconds resolution in our light curves
before becoming dominated by Poisson noise.  Despite this the light
curve data we do have, and the PSD of the 50 ks observation, are good
enough to tell us that NGC 5204 X-1 displays no strong variability on
timescales less than a few thousand seconds.  On timescales longer
than this NGC 5204 X-1 is varying, suggestive of a red noise PSD with
its weak power at higher ($\ga 10^{-3}$ Hz) frequencies swamped by
statistical noise in the data.

\subsubsection{A comparison with GRS 1915+105}

One interesting means of interpreting our temporal data is via a
direct comparison with GRS 1915+105 (see Fender \& Belloni 2004 for a
review of the properties of this remarkable object).  This source is
the archetypal Galactic microquasar, and has spent much of its (to
date) 15-yr outburst radiating at luminosities at or above the
Eddington limit for its $\sim 14 M_{\odot}$ black hole (Done,
Wardzi{\'n}ski \& Gierli{\'n}ski 2004).  This would make it a ULX to
an observer external to our Galaxy, hence it is potentially crucial
for our understanding of ULXs as a whole.  As it has a low-mass
companion, King (2002) suggested that sources akin to GRS 1915+105
could explain the lower-luminosity ULXs associated with old stellar
populations (such as the population of $1 - 2 \times 10^{39} \ergsec$
ULXs associated with elliptical galaxies; Irwin, Bregman \& Athey
2004).  A direct comparison between GRS 1915+105 and the brighter ULXs
- thought to have much more massive companion stars - may therefore
seem illogical.  However there may be some crucial similarities, for
example both types of system are thought to transfer mass through
Roche lobe overflow, and possibly accrete at super-Eddington rates.
Hence GRS 1915+105 may still be useful in interpreting the behaviour
of brighter ULXs.  For instance Goad et al. (2006) find that Ho II
X-1, which is far more luminous than GRS 1915+105 (at $> 10^{40}
\ergsec$) and a good IMBH candidate (Dewangen et al. 2005), shows no
temporal variability over short ($< 1$ hour) timescales during an
$\sim 80$ ks observation.  Interestingly, its light curve and PSD
during this observation appear to mirror the $\chi$-class of
behaviours of GRS1915+105 (Belloni et al. 2000).  Taken together with
a possible mass limit derived from PSD analysis, Goad et al. argue
that Ho II X-1 is a $< 100 M_{\odot}$ black hole accreting at very
high rates, and thus behaving similarly to GRS 1915+105 in its
``$\chi$'' (or ``plateau'') state.

One simple test is to qualitatively compare the long-term light curve
of GRS 1915+105 obtained from \rxte all-sky monitor (ASM) data
(Fig. 2, Fender \& Belloni 2004) to our monitoring light curve.
Despite a moderate bandpass difference (0.5 -- 8 keV for \chan ACIS-S
versus 1.5 -- 12 keV for the ASM), both light curves show a similar
amplitude variability, factors $\sim 5$ over a timescale of days.
Indeed, it is plausible that if one randomly sampled the GRS 1915+105
light curve at the same rate as our NGC 5204 X-1 light curve in
Fig.~\ref{fulllc}, one would find a very similar light curve.
However, this similarity is not necessarily maintained over shorter
timescales.  In particular GRS 1915+105 spends a reasonable fraction
of its time ($\sim 33 - 50\%$)\footnote{We calculate this from Table 1
of Belloni et al. (2000).  We assume that the 349 observation
intervals between January 1996 and December 1997 are representative of
the behaviour of GRS 1915+105, and that the $\chi$ and $\phi$ states
(that are present in $\sim 50\%$ of observations) will appear
invariant in our observations.  We note that the $\alpha, \gamma$ and
$\delta$ states (a further $\sim 17\%$) may also appear as such in our
comparatively low quality data.} displaying fairly extreme short-term
variability, characterised by complex, sometimes very repetitive
variability patterns in its light curves with cycle times ranging from
tens to thousands of seconds.  These are thought to be due to
limit-cycle instabilities in the inner accretion disc such that it is
continually emptying and refilling (Belloni et al. 2000)\footnote{In
this case the limit-cycle instability is thought to be driven by
cyclical changes in the balance between gas pressure and radiation
pressure.  This is distinct from a second form of limit-cycle
instability, namely the ionised versus neutral hydrogen cycle, which
dominates the long-term transient behaviour (i.e. outburst versus
quiescense) in accretion discs.}.  No such behaviour is seen from NGC
5204 X-1, even though we are sensitive to variations on timescales
$\la 100$ seconds in our data.

What does this imply about the nature of NGC 5204 X-1?  An important
clue could be that GRS 1915+105 only displays its unique limit-cycle
variability when it is accreting at super-Eddington rates (Done et
al. 2004).  This implies that this behaviour could be an observational
signature of super-Eddington accretion.  If so, the absence of such
behaviour in NGC 5204 X-1 would imply it is accreting at a
sub-Eddington rate.  In this case, we must either be observing beamed
X-ray emission (likely of the variety suggested by King et al. 2001)
or observing isotropic emission from a larger black hole, with the
peak luminosities of $6.3 \times 10^{39} \ergsec$ implying a black
hole mass in excess of $45 M_{\odot}$.  Interestingly, as most ULXs
also show little or no short-term variability (cf. Swartz et al. 2004,
Feng \& Kaaret 2005), and certainly nothing resembling the limit-cycle
variability of GRS 1915+105, this argument could imply that the vast
majority of ULXs are sub-Eddington accretors.

Of course, this is reliant on the limit-cycle behaviour of GRS
1915+105 being typical of all sources accreting above the Eddington
rate, and this is by no means assured.  For example, there are several
neutron stars that exceed the Eddington limit in our own Galaxy
(including Cir X-1 at up to 10 $L_{\rm Edd}$), none of which show
limit-cycle variability similar to GRS 1915+105, though Done et
al. (2004) note that irradiation of the accretion flow by the boundary
layer on the neutron star's surface might act as a stabilizing
mechanism.  Perhaps more pertinently, other Galactic black holes have
reached super-Eddington states (McClintock \& Remillard 2005), with no
reports of similar limit-cycle variability.  If we are to use the
apparent absence of limit-cycle variability as an argument that ULXs
are sub-Eddington accretors, then the key observation would be to
catch a ULX behaving in precisely this way when it radiates at its
brightest X-ray luminosities.  Otherwise, the lack of such behaviour
in ULXs might just serve to emphasize how unusual GRS 1915+105
actually is.

\subsection{On the soft part of the X-ray spectrum}

As Fig.~\ref{plfit} demonstrates, the soft end of our X-ray spectrum
showed unusual residuals, most notably a line-like feature peaking at
$\sim 1$ keV.  Our spectral modelling showed that this feature could
be described in two ways; either it is the most prominent feature
deriving from the presence of a warm ($kT \sim 0.9$ keV)
\textsc{mekal} optically thin thermal plasma, or it could be described
by a single, broadened ($\sigma \sim 70$ eV) Gaussian line at 0.96
keV, with a relatively low equivalent width ($\sim 50 - 80$ eV).
Whilst the other components of the spectral modelling were required to
vary to maintain good spectral fits throughout our campaign, the
contribution of this component remained the same.

So what is the origin of this feature?  ULXs in which a \textsc{mekal}
component models some part of their X-ray spectrum are relatively
scarce (though see e.g. Terashima \& Wilson 2004; Roberts et al. 2004;
Dewangen et al. 2005; Feng \& Kaaret 2005; Soria et al. 2006), and in
some of these cases we are faced with a very soft component that is
spectrally indistinguishable from a very soft accretion disc with the
quality of data available.  That is not the case in this observation,
where an optically-thin plasma clearly provides the best explanation
for the soft X-ray spectrum.  This fit is driven by the feature at
$\sim 1$ keV, similarly to the \textsc{mekal} component found to be
required to fit the spectrum of NGC 4395 X-1 (SRW06; Feng \& Kaaret
2005).

Feng \& Kaaret (2005) argue that a soft thermal component could arise
in a supernova remnant (SNR) co-located with the ULX.  For this to be
true, the contribution of the SNR to the X-ray flux of the ULX must
always be less than (or at most equal to) the total observed flux, as
a structure of light years in size cannot vary its X-ray flux on the
timescale of hours to days.  By examining the contribution of the
\textsc{mekal} components in Models (1) \& (2) to the fitted spectrum
of the 50-ks observation, we can infer that the \textsc{mekal} from
Model (1) contributes $\sim 0.03 \ctsec$ to the overall count rate,
which is plausible given the minimum observed count rates of $\sim 0.1
\ctsec$ for NGC 5204 X-1.  However, in Model (2) the \textsc{mekal} is
responsible for $\sim 0.16 \ctsec$, meaning it cannot originate in a
large structure like a SNR.\footnote{As the necessity to vary this
\textsc{mekal} component conflicts with every other model showing that
the soft component remains invariant as the ULX flux changes, we do
not consider this model as viable for explaining the spectral changes,
and so exclude it from the analysis in Section 4.3.}  If the thermal
component in Model (1) does originate in a SNR, we note that it cannot
be very old - its temperature of $\sim 0.96$ keV is rather similar to
the temperatures of the reverse shock components that become visible
on the timescale of years -- a couple of decades in recent supernovae
(e.g. Immler \& Lewin 2003) but somewhat higher than the $\sim 0.2 -
0.4$ keV temperatures of the fainter and older SNR population of M33
(Ghavamian et al. 2006).  The fact that the X-ray flux of this
component is seen to brighten by a factor $\sim 5$ in the eight months
between the \xmm observation and the \chan programme (during which it
remains constant) may be difficult to reconcile with a large structure
such as an SNR, though we note that the X-ray emission of SN 1987A has
been seen to evolve relatively rapidly (on the timescale of months) as
its blast wave propagates outwards and hits dense clumps of ISM,
causing X-ray ``hotspots'' to turn on (e.g. Park et al. 2004).
However, in this case it is puzzling that one does not detect any
bright radio emission from NGC 5204 X-1 (K{\"o}rding, Colbert \&
Falcke 2005), as most X-ray bright recent supernovae also appear radio
bright (Immler \& Lewin 2003).

If the feature does not arise in an underlying SNR, where else could
it originate?  If we consider the Gaussian line fit, its line width of
$\sim$ 60-80 eV corresponds to velocities of 0.06-0.08$c$ (18000-24000
$\kmsec$) for matter radiating a single emission line.  Whilst these
velocities may be plausible for the innermost regions of an accretion
disc and/or a mildly relativistic outflow, in the absence of direct
evidence for a relativistic origin (for example, a classic skewed line
profile, or a measurement of its blue- or redshift), or indeed a good
candidate for a single dominant emission line at this energy, perhaps
a more realistic interpretation is that the feature originates as a
blend of the Ne K and Fe L lines prominent around 0.96 keV.

It is once again pertinent to raise GRS1915+105 at this point, as
recent \xmm EPIC observations have shown that it displays a remarkably
similar spectral feature, namely a broad ($\sigma \sim 90$ eV)
emission line at 0.97 keV, albeit with a much higher equivalent width
of 5.6 keV (Martocchia et al. 2006).  Whilst there are caveats
attached to this detection - most notably the non-detection of a
corresponding feature in simultaneous \xmm RGS data - this finding
again raises the question of whether there are at least some physical
similarities between GRS1915+105 and NGC 5204 X-1.  Martocchia et
al. (2006) interpret the line feature as possible evidence for a
photoionized disc wind (a collisionally-ionized wind is also
discussed, but rejected because calculations show any such wind in GRS
1915+105 must be approaching the optically-thick regime).  The same
interpretation could apply to NGC 5204 X-1, though the lower
equivalent width of its line requires an explanation (speculations
could include the continuum level being higher relative to the wind
emission, or the wind emission simply being less bright in NGC 5204
X-1 than GRS 1915+105).  If so, this could be the first evidence of a
photoionized disc wind contributing to the X-ray emission of a ULX.

\subsection{On the overall spectral shape and its variation}

In Table~\ref{specfits} we show five spectral models, four of which
are broadly empirical and the fifth more physically-based, which can
adequately describe the 50-ks \chan ACIS-S spectrum of NGC 5204 X-1.
We can arguably rule one fit (number 2) out.  We are then left with
models that (if taken at face value) variously describe (1) a compact
non-thermal emitter surrounded by a near-solar metallicity thermal
plasma, (3) a fairly ``standard'' representation of an accretion disc
plus corona spectrum with a broad emission line just short of 1 keV,
(4) a standard accretion disc plus soft thermal excess model with the
same emission line and (5) a physical accretion disc plus corona
model, again with the broad emission line.  Unfortunately even with
the relatively good quality of a $\sim 9000$ count ACIS-S spectrum we
can only tell that these are all adequate descriptions of the spectral
shape, and not begin to distinguish between these models in a
statistical sense.  We are then left to make the choice as to whether
we draw our physical insight from the simpler, empirical models by
comparison to the fits these models make to better understood sources,
or whether we trust the insight of more physical models at the
possible cost of allowing more degrees of freedom and/or potential
ambiguity if the best-fitting parameters show large uncertainty
ranges.  Both methods have benefits and drawbacks (with opposite
viewpoints outlined by Miller, Fabian \& Miller 2006 and Goncalves \&
Soria 2006).  Here, we will consider the insights provided by the
physical model further.

SRW06 find that the majority of ULXs with high-quality \xmm EPIC
spectral data display a spectral break (which could also be described
as a cut-off, or interpreted as an indicator of curvature) above 2
keV, which could be an important clue as to the physical nature of
ULXs.  These authors are not the only ones to identify such a feature
in ULXs; this was firstly seen by Kubota, Done \& Makishima (2002) in
\asca GIS data from IC 342 X-1, and more recently has been seen for
other ULXs observed by {\it XMM-Newton\/}, notably the top IMBH
candidates Ho IX X-1 (Dewangen, Griffiths \& Rao 2006) and M82 X-1
(Agrawal \& Misra 2006).  This break argues against the simple IMBH
model where the emission above 2 keV is dominated by a single
power-law continuum representative of an optically-thin hot corona (as
observed in most Galactic black hole X-ray binaries), and could
indicate the presence of unusual physical processes in the system.
One such explanation discussed by SRW06 is the presence of an
optically-thick corona.  Though such fits require the seeding of the
corona by a disc with an IMBH-like temperature, SRW06 refer to two
models (Zhang et al. 2000; Done \& Kubota 2006) viably describing very
high accretion rate systems - with no requirement for an IMBH - that
could explain both the apparently cool disc emission, and the
optically thick corona (see also Kuncic \& Bicknell 2004 for a similar
model).

{\it Our physical model (5) describes the continuum emission of NGC
5204 X-1 as an optically thick corona seeded by a cool disc\/},
consistent with the modelling of other ULXs from \xmm
data.\footnote{Obviously we have a second, line-like component in our
spectral fit.  For the purposes of this discussion we treat its origin
as separate from the underlying continuum.}  Importantly, the \chan
data is of sufficient quality to demonstrate that this solution is
preferred at the $3\sigma$ level to an optically-thin solution (with
the latter also taking an implausibly low disc temperature of 20 eV).
Perversely though, the \chan data alone is not of good enough quality
to rule out an unbroken power-law above 2 keV, with the simple
power-law versus broken power-law test of SRW06 inconclusive in this
instance.  This is perhaps unsurprising given that there are only
three data points above 5 keV in the \chan data, and that models
including a hard power-law [(1), (3)] do provide good fits to the
data.  In this case we can turn to the \xmm data for this ULX for
confirmation, which as SRW06 demonstrate does indeed show a
significant break feature at $\sim 5$ keV.

However, if we are to maintain any faith in this physical model, its
parameters must vary in a physical way throughout the monitoring
programme.  Our modelling of data taken from within the 50-ks
observation, and from the full monitoring programme, show that the
hardening of the spectra with increasing flux (graphically
demonstrated by the hardness ratios of Fig.~\ref{hrs}) can be modelled
predominantly as an increase in the temperature of the corona as the
ULX luminosity increases, with additional changes in the seed disc
temperature occuring over timescales of days.  This trend is not just
seen in the \chan data; as we have shown, an \xmm observation is also
consistent with this behaviour.  Interestingly, this echoes the
behaviour of XTE J1550-564 in a subset of its very high state that
Kubota \& Done (2004) describe as a ``strong very high state''.  This
state is characterised by strong Comptonisation from a cool (albeit in
this case still 10 - 30 keV) corona, and unusually low disc
temperatures for its luminosity ($\sim 0.5$ keV against an expectation
of $\sim 1$ keV for that luminosity in the high/soft state).  In this
state the temperature of the corona does indeed seem to track the
luminosity of the system (see Fig. 5 of Kubota \& Done 2004).  We
therefore speculate that NGC 5204 X-1 is in a more extreme version of
this strong very high state, characterised by a cooler disc and a
thicker corona.  We further speculate that this thicker corona could
be fueled by the extreme mass transfer rates possible for ULXs if they
are stellar-mass black holes accreting from young, massive stars as
described by Rappaport, Podsiadlowski \& Phahl (2005).  Indeed, Done
\& Kubota (2006) note that this state has the characteristics to
explain the X-ray spectra of many ULXs without the need to resort to
IMBHs.

We therefore conclude that the optically-thick solution remains a
physically plausible model for this ULX, though we caution that it is
by no means an unique solution for this data.  As reported in previous
work, this could again mean that we are looking at a very high
accretion rate stellar-mass (or slightly larger) black hole.  In this
state it could share some similarities with GRS 1915+105, though we
have shown that limit-cycle variability does not appear to be one of
them.  A possible outflow identified by the soft, broad emission
line-like component identified at $\sim 0.96$ keV is however one
potential similarity.  Whilst the possible detection of an optically
thick corona continues a current trend for bright ULXs (and should not
be a surprise given SRW06 detected a similar spectrum in \xmm data for
this ULX), the detection of the broad line feature is unexpected.  The
presence of this line is certainly very unusual when compared to the
range of ULX characteristics reported to date, and as such it should
be fully investigated by future studies, in particular by utilising
the reflection grating spectrometers on \xmm to probe the composition
of this feature.

\section{Conclusion}

We have reported the results of a two-month monitoring programme of
NGC 5204 X-1, over timescales of days to weeks.  We detect no periodic
variability, and the source shows little variability at all on
timescales less than an hour, though it varies by factors of $\sim 5$
over a few days.  This flux variation is accompanied by spectral
changes, in the sense that the source spectrum becomes harder as the
flux increases.  The spectrum of the ULX is composed of two main
features: a broad emission line at $\sim 0.96$ keV, and a smooth
underlying continuum, that can together be described by a variety of
multi-component models.  One of these models describes a continuum
originating in an optically-thick corona, consistent with other recent
results for ULXs, that could describe an extreme form of a strongly
Comptonised very high state that is seen in Galactic black hole X-ray
binaries.  The spectral variations appear consistent with the known
behaviour of sources in this state.  This model does not require an
IMBH to power the ULX.  Of course, this is not the same as ruling out
the presence of an IMBH in this system.

The insights gained from studying variations in the X-ray spectra and
temporal characteristics of Galactic black holes over timescales from
milliseconds to years have been immense, and have been facilitated by
the unique capabilities of the \rxte satellite (McClintock \&
Remillard 2006 and references therein).  Such studies cannot currently
be done for any single extragalactic ULX, with the exception of M82
X-1 (Kaaret, Simet \& Lang 2006a,b), and even that study is limited by
effects such as the possible confusion of multiple ULXs in the centre
of M82.  For the remaining fainter sources we are currently
constrained to small monitoring campaigns using \xmm and {\it
Chandra\/}, which have been surprisingly rare to date given the
potential physical insights they can offer.  Future, sensitive all-sky
monitors with arcminute resolution such as {\it Lobster\/} (Fraser et
al. 2002) will at least offer a reasonable monitoring platform for the
brightest ULXs so that we can finally examine the details of their
day-to-day behaviour.  Until then, improvement in our insights into
the physical processes of ULXs - and hence their underlying nature -
will continue to rely on obtaining deeper single observations, or more
monitoring of the type described in this paper.

\vspace{0.2cm}

{\noindent \bf ACKNOWLEDGMENTS}\\ We would like to thank an anonymous
referee for helping us to improve this paper.  TPR gratefully
acknowledges support from the UK Particle Physics and Astronomy
Research Council (PPARC).  The authors would also like to thank Simon
Vaughan for his advice on temporal analyses, and Chris Done for
discussions on Galactic black holes.  This work is in part based on
observations obtained with {\it XMM-Newton\/}, an ESA science mission
with instruments and contributions directly funded by ESA member
states and the USA (NASA).

\label{lastpage}

\end{document}